\documentclass{article}
\usepackage{fullpage}
\usepackage{amssymb,latexsym,amsmath}     
\usepackage{graphicx}
\usepackage{authblk}
\usepackage{dsfont}
\usepackage{amsthm}
\usepackage{color}
\usepackage{float}
\usepackage{hyperref}
\usepackage{subfigure}
\usepackage{float}
\usepackage{multirow}
\usepackage{graphicx} 
\usepackage{amsmath}
\usepackage{amssymb}
\usepackage{calc}
\usepackage{mathtools}
\usepackage{tcolorbox}
\usepackage[mathscr]{euscript}
\usepackage{cuted}
\usepackage{quiver}
\usepackage{amsmath}
\usepackage{tikz}
\usetikzlibrary{arrows.meta,decorations.pathreplacing,calc}

\title{Computing nonlinearity ratios in black hole ringdown}
\begin{document}
\begin{center}
\vspace{24pt} { \large \bf Computing nonlinearity ratios using second order black hole perturbation theory } \\
\vspace{30pt}
\vspace{30pt}
\vspace{30pt}
{\bf Jasveer Singh\footnote{jasveer.singh@students.iiserpune.ac.in}}, {\bf Vardarajan
Suneeta* \footnote{suneeta@iiserpune.ac.in}}\\
\vspace{24pt} 
{\em * The Indian Institute of Science Education and Research (IISER),\\
Pune, India - 411008.}
\end{center}
\date{\today}
\bigskip
\begin{center}
{\bf Abstract \\ }
\end{center}
We revisit an analytical approximation scheme for computing nonlinearity ratios involving quadratic quasinormal modes (QQNMs). We compute these ratios for the general case when the QQNM is not one of the linear QNMs, for the $(l,m)$ channel $(2,2) \times (2,2) \to (4,4)$. We find an excellent match with numerical simulations. We also discuss where and why the method can fail, for example, for the channel $(2,0) \times (2,0) \to (2,0)$, where we can only get crude estimates for the nonlinearity ratio. Motivated by recent studies on nonlinear ringdown at the horizon, we also compute the nonlinearity ratios at the horizon. We find that the ratio both at the horizon and infinity is insensitive to different choices of regularization of the source term in the second order perturbations. We also discuss amplitudes of QQNMs sourced by linear overtones. Finally, we discuss the issues that must be resolved within this method to do precision analysis of nonlinear ringdown.

\newpage
\section{Introduction}
In binary black hole mergers, black hole linear perturbation theory and quasinormal modes (QNMs) have been used to describe the ringdown phase. Recently, nonlinear effects during ringdown have been probed in simulations and the importance of studying \textit{second} order perturbations has been recognized \cite{Cheung_2023}, \cite{Mitman_2023}, \cite{Lagos_2023}.\\

Perturbation theory beyond linear order around a Schwarzschild black hole was studied a long while ago in \cite{Gleiser2_1996}\cite{Nicasio_1999}, \cite{Gleiser_2000}, \cite{Brizuela_2006}, \cite{Brizuela_2007}, \cite{Nakano_2007}, \cite{Brizuela_2009} (see also \cite{Brizuela_2010} for perturbations of a star).
A useful summary of these results can be found in \cite{spiers2024secondorderperturbationsschwarzschildspacetime}.
With an appropriate choice of gauge, the linearized Einstein equations for perturbations of the Schwarzschild geometry imply the Regge-Wheeler \cite{PhysRev.108.1063} and Zerilli \cite{PhysRevLett.24.737} equations for scalars formed from the perturbations.
The causal Green's function for the linearized operator has an infinite but discrete set of complex frequency poles $\omega_{l,m,n}$, the QNM frequencies, and the gravitational perturbations at intermediate times during the ringdown phase can be described by a sum of exponentially damped QNMs. These frequencies are characterized by two angular harmonic numbers $(l,m)$ and an overtone number $n$. The real and imaginary part of $\omega_{l,m,n}$ thus determine the QNM oscillation frequency and decay timescale respectively. The gravitational wave strain amplitudes as $r \to \infty$ are denoted by $A_{l,m,n}$ such that, as $r \to \infty$,
\begin{eqnarray}
r\,h_{l,\,m}(T-r)&=&\sum_{n\geq 0}A_{l,m,n} e^{-i\omega_{l,m,n}(T-r)}.
\end{eqnarray}
Here, $T$ is the usual Schwarzschild time coordinate.

Second-order QNMs are those QNMs that obey either the Regge-Wheeler or Zerilli equations, with a source term quadratic in the first-order QNMs. The source term is obtained by studying the Einstein equation in weakly nonlinear perturbation theory.
It was pointed out, for example, in \cite{Nakano_2007}, that the dominant second order QNMs sourced by the dominant first order QNMs can have an amplitude of up to $10 \%$ of the amplitude of the dominant first order QNM and can be more significant than linear overtones.
The quadratic quasi-normal modes (QQNMs) are generated by the product of two linear modes through the channel $(l,m) \times (l',m') \rightarrow (L,M)$, subject to standard angular momentum selection rules. These non-linear effects are thus generically quantified by the ratio of the amplitudes of the second-order non-linear gravitational wave strain $A_{l,m,n \times l',m',n'}$ with frequency
\begin{equation}
\omega_{{l_1,m_1,n_1\times l_2,m_2,n_2}}=\omega_{l_1,m_1,n_1}+\omega_{l_2,m_2,n_2},
\end{equation}
to the product of the amplitudes of its parent modes
\begin{equation}
{\rm NL}_h =\left|\frac{A_{l_1,m_1,n_1\times l_2,m_2,n_2}}{A_{l_1,m_1,n_1}A_{l_2,m_2,n_2} }\right|. \label{ratio}
\end{equation}
Dominant linear QNMs are $(2, \pm 2)$ \cite{Berti_2007} in quasicircular mergers,
and $(2,0)$ for head-on collisions.
Recently, the nonlinearity ratio for the most excited non-linear QQNM $(2,2)\times(2,2)\rightarrow(4,4)$ has been obtained from binary merger simulations in numerical relativity \cite{Cheung_2023},\cite{Mitman_2023}, second order perturbation theory and numerics in \cite{Redondo_Yuste_2024} and by applying the Leaver algorithm \cite{Bucciotti_2024}. All obtain a similar value for the nonlinearity ratio at infinity
\begin{eqnarray}
    NL_{2,2,0 \times 2,2,0} = \left|\frac{A_{2,2,0\times2,2,0}}{A^2_{2,2,0}}\right|\sim 0.15-0.20.
\end{eqnarray}
The reason why this is a range, rather than one number is explained and shown in \cite{bourg2025quadraticquasinormalmodedependence}. In the Schwarzschild spacetime, both even and odd parity modes contribute to the same QQNM due to isospectrality. But different processes excite different amplitudes for the even and odd sector, leading to a range for the nonlinearity ratio.
Nonlinearity ratios have also been computed numerically in the eikonal limit in \cite{JRY}.
Further, in a recent work \cite{Khera_2023}, the nonlinearity ratios have been calculated for the first time at the black hole horizon.\\

The nonlinearity ratio has been computed analytically by Perrone, Barreira, Kehagias and Riotto (PBKR)\cite{Perrone_2024}, in a WKB approximation, using the method of steepest descent. More recently, it has been computed using other methods in \cite{kehagias2024nonlineareffectsblackhole}, in the eikonal limit in \cite{kehagias2025nonlinearquasinormalmodesschwarzschild} and in \cite{kehagias2025nonlinearitiesschwarzschildblackhole}.
The nonlinearity ratio taking into account both even and odd modes have been computed in \cite{bourg2025quadraticquasinormalmodedependence}.
In this paper, we revisit the computation of PBKR \cite{Perrone_2024}, and discuss some of the issues involved in the computation of the nonlinearity ratios using the WKB approximation, matched asymptotic expansions and the method of steepest descent. As we discuss in section \ref{sec8}, the case studied in \cite{Perrone_2024} corresponds to studying QQNMs whose frequency is also one of the linear QNM frequencies. This is not the case in general, for QNMs of the Schwarzschild spacetime. We therefore study the case where the QQNM frequency, which is the sum of linear QNM frequencies, is not itself a linear QNM frequency.

We compute the nonlinearity ratio for QQNMs sourced by the dominant linear modes $(l,m) = (2, \pm 2)$ in the analytical approximation scheme of PBKR \cite{Perrone_2024}. Specifically, we apply the WKB approximation \cite{1985ApJ...291L..33S},\cite{Perrone_2024} and matched asymptotic expansions to get approximate solutions for the QQNMs. We then calculate the nonlinearity ratios in this approximation scheme using the method of steepest descent. As mentioned before, since we assume that the QQNM frequencies are not themselves linear QNM frequencies of the system, this leads to different matching conditions from PBKR. We compute nonlinearity ratios both at infinity and at the black hole horizon, and for parent modes being overtones as well.
For the channel $(2,2) \times (2,2) \to (4,4)$, this method yields a nonlinearity ratio that agrees very well with that obtained in numerical simulations \cite{Cheung_2023}. For the channel $(2,0) \times (2,0) \to (2,0) $, on the other hand, this method gives a too large nonlinearity ratio. We give evidence that this is due to the fact that in this case, the method of steepest descent fails. Further, in this case, it is not possible to get a \textit{precise} nonlinearity ratio by integrating the relevant integral numerically, since the QQNMs are sourced by \textit{spatially truncated} linear QNMs whose \textit{exact} support is not known \cite{szpak2004quasinormalmodeexpansionexact}, \cite{Okuzumi:2008}. We can only get a rough estimate using numerical integration. Recently, the nonlinearity ratio for this channel has been computed using different methods in \cite{kehagias2025nonlinearitiesschwarzschildblackhole} by using the nonlinearity ratio for the channel producing $l=4$.

The paper is organized as follows. In section \ref{sec2} we give a general review of how the first and second-order wave equations are set up and obtained using second-order black hole perturbation theory. In section \ref{sec3}, we review the Zerilli equations at first and second order. In sections \ref{sec4} and \ref{sec5}, following \cite{Perrone_2024}, we review the solution to the Zerilli equations using the Green's function method, to obtain the exact first and second-order solutions respectively. In section \ref{sec6} we review the WKB approximation and the quasinormal mode solutions. We then obtain the general homogeneous solutions in section \ref{sec7} and section \ref{sec8} in the situation when the frequency is not a linear QNM frequency. Using these solutions, we calculate the nonlinearity ratio for the $l=2,m=0$ QQNM at infinity and the black hole horizon in section \ref{sec9} and discuss why the analytical approximation used (method of steepest descent) fails in this case. We discuss the issues involved in a \textit{precise} computation of the nonlinearity ratio for this channel. We compute the nonlinearity ratio for the $l=4,m=4$ QQNM in section \ref{sec10}, and obtain excellent agreement to numerical results. We also discuss the (in)sensitivity of this ratio to the choice of matching points, and the choice of regularization of source term, in the same section. We then discuss the nonlinearity ratios for QQNMs excited by linear overtones in section \ref{sec11}. We finally summarize our results and discuss them in section \ref{sec12}.

\section{Setting up the equations: review}\label{sec2}
We now summarize second order perturbation theory.
Let us consider second order metric perturbations of the Schwarzschild metric
\begin{equation}
g_{\mu\nu} = g^{(0)}_{\mu\nu} + h^{(1)}_{\mu\nu}+h^{(2)}_{\mu\nu},
\label{so1}
\end{equation}
where the Schwarzschild metric in $(T,r, \theta, \phi)$ coordinates is given as
\begin{equation}
    g^{(0)}_{\mu\nu}dx^{\mu}dx^{\nu} = -\left(1-\frac{2M}{r}\right)dT^2 +\frac{1}{\left(1-\frac{2M}{r}\right)}dr^2 +r^2(d\theta^2+sin^2\theta d\phi^2) .
\label{so2}
\end{equation}

In the perturbative expansion \eqref{so1}, it is understood that there is a small parameter $\epsilon$, such that $| h^{(1)}_{\mu\nu}| \sim O(\epsilon)$ and
$| h^{(2)}_{\mu\nu}| \sim O(\epsilon^{2})$.
Now, for perturbations about the Schwarzschild metric up to second order, we can calculate the Einstein tensor up to second order (see, for example, \cite{Nakano_2007}). Omitting the spacetime indices $\mu$ ,$\nu$ of the metric perturbations $h^{(1)}_{\mu\nu}$ and $h^{(2)}_{\mu\nu}$ for clarity and neglecting terms of third and higher perturbative order gives us
\begin{equation}
    G_{\mu\nu}[g_{\mu\nu}] =
    G_{\mu\nu}^{(1)}[h^{(1)}]+G_{\mu\nu}^{(1)}[h^{(2)}]+G_{\mu\nu}^{(2)}[h^{(1)},h^{(1)}]
+O((h^{(1)})^3,\,h^{(1)}h^{(2)},\,(h^{(2)})^2).
\label{so3}
\end{equation}
$G_{\mu\nu}^{(1)}[h]$ here is the linearized Einstein tensor,
\begin{eqnarray}
G_{\mu\nu}^{(1)}[h] &=&
-\frac{1}{2}h_{\mu\nu;\alpha}{}^{;\alpha}+B_{(\mu;\nu)}
-R_{\alpha\mu\beta\nu}h^{\alpha\beta}
-\frac{1}{2}h_{;\mu\nu} -\frac{1}{2} g_{\mu\nu}
(B_{\lambda}{}^{;\lambda}-h_{;\lambda}{}^{;\lambda}) \,; \nonumber \\
B_{\mu} &=& h_{\mu\alpha}{}^{;\alpha} \;.
\label{so4}
\end{eqnarray}
$h_{\mu\nu}$ here stands for $h^{(1)}_{\mu\nu}$ or $h^{(2)}_{\mu\nu}$. Further, the semi-colon $;$ denotes a covariant derivative taken with respect to the background Schwarzschild metric $g_{\mu\nu}^{(0)}$ and raising and lowering of indices is also done with respect to this metric.

$G_{\mu\nu}^{(2)}$ would thus have terms that are quadratic in the first-order perturbation
\begin{eqnarray}
G_{\mu\nu}^{(2)}[h^{(1)},h^{(1)}] &=&
R_{\mu\nu}^{(2)}[h^{(1)},h^{(1)}] - \frac{1}{2}g_{\mu\nu}R^{(2)}[h^{(1)},h^{(1)}] -\frac{1}{2} h_{\mu\nu}^{(1)} R^{(1)}[h^{(1)}]
\,; \nonumber \\
R_{\mu\nu}^{(2)}[h^{(1)},h^{(1)}]
&=& \frac{1}{4}h^{(1)}_{\alpha\beta;\mu}h^{(1)}{}^{\alpha\beta}{}_{;\nu}
+\frac{1}{2}h^{(1)}{}^{\alpha\beta}(h^{(1)}_{\alpha\beta;\mu\nu}
+h^{(1)}_{\mu\nu;\alpha\beta}
-2h^{(1)}_{\alpha(\mu;\nu)\beta}) \nonumber \\
&&
-\frac{1}{2}(h^{(1)}{}^{\alpha\beta}{}_{;\beta}-\frac{1}{2}h^{(1)}_{\beta}{}^{\beta;\alpha})
(2h^{(1)}_{\alpha(\mu;\nu)}-h^{(1)}_{\mu\nu;\alpha})
+\frac{1}{2}h^{(1)}_{\mu\alpha;\beta}h^{(1)}_{\nu}{}^{\alpha;\beta}
-\frac{1}{2}h^{(1)}_{\mu\alpha;\beta}h^{(1)}_{\nu}{}^{\beta;\alpha} \,.
\label{so5}
\end{eqnarray}
Now that we have the complete Einstein tensor up to second perturbative order, we can put this into the Einstein equation to solve and obtain $h_{\mu\nu}^{(1)}$ and  $h_{\mu\nu}^{(2)}$.

We consider the vacuum Einstein equation.
We get
\begin{equation}
    G_{\mu\nu}[g_{\mu\nu}] =
    G_{\mu\nu}^{(1)}[h^{(1)}]+G_{\mu\nu}^{(1)}[h^{(2)}]+G_{\mu\nu}^{(2)}[h^{(1)},h^{(1)}]
    =0.
\label{so6}
\end{equation}
We set the left-hand side of \eqref{so6} to zero order by order in $\epsilon$.
\begin{equation}
    G_{\mu\nu}^{(1)}[h^{(1)}] = 0,
\label{so7}
\end{equation}
at the first order, and
\begin{eqnarray}
        &G_{\mu\nu}^{(1)}[h^{(2)}]+G_{\mu\nu}^{(2)}[h^{(1)},h^{(1)}] = 0 \nonumber \\
        \implies &G_{\mu\nu}^{(1)}[h^{(2)}] = - G_{\mu\nu}^{(2)}[h^{(1)},h^{(1)}]. \label{so8}
\end{eqnarray}
at the second order.

We first solve \eqref{so7} for $h^{(1)}$, and substitute that into the right-hand side of \eqref{so8}.
We see that \eqref{so8} is similar to the first-order equation, but now with a source term - an effective stress-energy tensor that is made out of the linear perturbations.

\section{The Zerilli equation}\label{sec3}
We will now follow the approach of Regge and Wheeler \cite{PhysRev.108.1063} and Zerilli \cite{PhysRevLett.24.737}.
We have to solve $7$ equations for the even perturbations and $3$ equations for the odd perturbations. This proves to be quite tiresome and we shall thus construct certain scalars out of these tensors and instead solve for them. After obtaining the scalars, we can reconstruct the metric perturbation terms from them.

In numerical simulations of merging binaries, it is seen that the dominant modes excited are the even ones. In this paper, we will consider as a seed linear perturbation, two cases: the dominant even linear quasinormal mode $l=2, m=2, n=0$ , and the dominant even mode in a head-on collision, $l=2,m=0, n=0$.

The full metric perturbation of the Schwarzschild black hole in a spherical harmonic decomposition can be found in \cite{Nakano_2007} (equation 3.1 of that paper).
Imposing the Regge-Wheeler gauge conditions and only considering the even parity modes, we can write the metric perturbations as the following matrix:
\begin{equation}
    {\textbf h}^i = \begin{bmatrix}
(1 -\frac{2M}{r})H^{(i)}_{0lm} & H^{(i)}_{1lm} & 0 & 0\\
H^{(i)}_{1lm} & (1 - \frac{2M}{r})^{-1}H^{(i)}_{2lm} & 0 & 0 \\
0 & 0 & r^2K^{(i)}_{lm} & 0\\
0 & 0 & 0 & r^2 sin^2\theta K^{(i)}_{lm}\\
\end{bmatrix}Y_{lm},
\label{so9}
\end{equation}
where $i=1,2$ refer to first order and second order perturbations respectively and $Y_{lm}$ are the scalar spherical harmonics. There is a summation over $l,m$.

We now introduce the scalars $\psi$ and $\chi$ to reduce the $7$ even wave equations at each order to a single scalar equation. As given in \cite{Nakano_2007}, we define
\begin{eqnarray}
\psi^{(1)}_{l m}(T,r)&=&
\frac{r}{\lambda+1}
\left[
K_{l m}^{(1)} (T,r)
+{\frac { r-2\,M  }
{  \lambda\,r+3\,M  }}
\left(H_{2\,l m}^{(1)} ( T,r )
-r\,
{\frac {\partial }{\partial r}}K_{l m}^{(1)}(T,r)\right)
\right]
\,;
\nonumber \\
\lambda &=& \frac{(l-1)(l+2)}{2} \,.
\label{so10}
\end{eqnarray}
This scalar $\psi_{lm}^{(1)}$ now obeys the first order Zerilli equation,
\begin{eqnarray}
\left[-\frac{\partial^2}{\partial T^2}
+\frac{\partial^2}{\partial r_*^2}
-V_{Z}(r)\right] \psi_{\ell m}^{(1)}(T,r) = 0, \label{so11}
\end{eqnarray}
where
\begin{equation}
    V_Z(r) = \left(1-\frac{2M}{r}\right)
\frac{2\lambda^2 (\lambda+1) r^3+6 \lambda^2 M r^2+
18 \lambda M^2 r +18 M^3}{r^3 (\lambda r+3M)^2}\label{so12}
\end{equation}
is the Zerilli potential, and
\begin{equation}
    r_* = r+2M \ln \left(\frac{r}{2M}-1\right) ,\label{so13}
\end{equation}
is the tortoise coordinate.

Similarly, we introduce the second-order perturbation scalar \cite{Nakano_2007}
\begin{eqnarray}
\chi^{(2)}_{lm}(T,r)=\frac{r-2M}{3 (3 r+M) }
\left[\frac{r^2}{r-2M}
\frac{\partial K_{lm}^{(2)}(T,r)}{\partial T}
- H_{1\,lm}^{(2)}(T,r)\right].
\label{eq:defchi}
\end{eqnarray}
This satisfies the second-order Zerilli equation with an effective source term quadratic in $\psi_{\ell m}^{(1)}(T,r)$;
\begin{eqnarray}
\left[-\frac{\partial^2}{\partial T^2}
+\frac{\partial^2}{\partial r_*^2}
-V_{Z}(r)\right] \chi^{(2)}_{lm}(T,r) &=& S_{lm}(T,r) \,.
\label{eq:Z2}
\end{eqnarray}
Thus solving \eqref{so11} gives us $\psi$, which we then use to calculate the second-order source term in \eqref{eq:Z2} --- we then solve the equation to obtain $\chi$. Finally, we deconstruct $\psi$ and $\chi$ to obtain back the metric perturbations $h_{\mu\nu}^{(1)}$ and $h_{\mu\nu}^{(2)}$.

\section{First-Order Solutions: review}\label{sec4}
We have to now solve the first and second order Zerilli equations \eqref{so11} and \eqref{eq:Z2}. In this section and the next, we follow PBKR \cite{Perrone_2024} and use their notation.
Consider the following initial conditions for first and second order perturbations,
\begin{equation}
    \psi(x,0) = f_1(x), \quad \dot{\psi}(x,0) = g_1(x),
\end{equation}
\begin{equation}
    \chi(x,0) = f_2(x), \quad \dot{\chi}(x,0) = g_2(x),
\end{equation}
along with the QNM boundary conditions for each mode with frequency $\omega$:
\begin{equation}
    \psi_{\omega}(x,T) \to e^{-i\omega x}e^{i\omega T}, \text{\quad   as  }x\to \infty,
\end{equation}
\begin{equation}
    \psi_{\omega}(x,T) \to e^{i\omega x}e^{i\omega T},  \text{\quad   as  }x\to -\infty.
\end{equation}
Following Nollert and Schmidt \cite{PhysRevD.45.2617},
we will take the Laplace transform of the perturbation rather than the Fourier transform.
The Laplace transform of a function of $T$, $f(T,x)$ is
\begin{equation}
    \mathcal{L}[f] =\tilde{f}(s)= \int_0^{\infty}{\rm d} t \,e^{-sT}\, f(T),
\end{equation}
and its inverse as
\begin{equation}
    \mathcal{L}^{-1}[\tilde{f}] =f(T)= \int_{\epsilon-i\infty}^{\epsilon + i\infty} {\rm d} s \, e^{sT} \,\tilde{f}(s).
\end{equation}
As in \cite{Perrone_2024}, we take a Laplace transform of \eqref{so11}, and use the QNM boundary conditions to get
\begin{equation}
   ( \partial^2_x - s^2 - V(x))\,\psi_1(x,s) = -s f_1(x) - g_1(x),
\label{gf}
\end{equation}
and
\begin{equation}
    \psi_1(x,s) \to e^{- s x}, \quad x\to \infty,
\end{equation}
\begin{equation}
    \psi_1(x,s) \to e^{s x}, \quad x\to -\infty\; ;
\end{equation}
where we can consider $s = i\omega$. Consider two linearly independent solutions to the homogeneous equation, with the right hand side of \ref{gf} set to zero, denoted by $\phi_+(x,s)$  and $\phi_-(x,s)$, such that $\phi_+(x,s)$ obeys the QNM boundary condition at $\infty$ and $\phi_-(x,s)$ obeys the QNM boundary condition at the horizon. The Green's function for the problem is
\begin{equation}
G(x,x',s) = \frac{\phi_+(x_>,s)\, \phi_-(x_<, s)}{W(s)},
\end{equation}
where
\begin{equation}
x_> = \max(x, x'), \quad x_< = \min(x, x'),
\end{equation}
and $W(s) = \phi_{-}(x,s)\frac{\partial}{\partial x} \phi_{+}(x,s) - \phi_{+}(x,s)\frac{\partial}{\partial x} \phi_{-}(x,s)$
is the Wronskian, which is independent of $x$.
We obtain the solution to \ref{gf},
\begin{equation}
\psi_1(x,s) = \int_{-\infty}^{+\infty} \; {\rm d}x'\;  \frac{\phi_+(x_>,s)\, \phi_-(x_<, s)}{W(s)} \left( -s\,f_1(x') - g_1(x') \right).
\label{lt}
\end{equation}
We now take an inverse Laplace transform of this solution. This is an integral in the complex $s$ plane, and the Green's function will have both poles (from the zeroes of the Wronskian), and a branch cut. We assume there are no other poles of the integrand in \ref{lt}. The branch cut leads to power law tails at late times, while the poles are at the QNM frequencies, and affect the perturbation at intermediate times. Since we are interested in the QNM behaviour of the perturbations, we ignore the branch cut.

The inverse Laplace transform picks up residues at the poles of the Green's function, which correspond to $\phi_{+}$ and $\phi_{-}$ being linearly dependent, and happens at special $s_n$ solving $W(s_n) = 0$. In the notation of \cite{Perrone_2024}, we get:
\begin{equation}
\psi_1(x,T) = \sum_{n} \;c_{n{\rm L}}\; \phi_{-}(x,s_n) \;e^{Ts_n} ,\label{psi}
\end{equation}
where
\begin{equation}
c_{n{\rm L}}=W'(s_n)^{-1}\; \int_{-\infty}^{+\infty} \; {\rm d}x'\; \phi_{-}(x', s_n) \left( -s_n\,f_1(x') - g_1(x') \right).\label{cnL}
\end{equation}
Thus, $c_{n{\rm L}}$ counts the contribution to the solution from the $n-$th pole of the Green's function.

When the Wronskian is zero, as we said, $\phi_-$ and $\phi_+$ are linearly dependent. This means that the solutions obey the boundary condition at both infinity and horizon and thus correspond to the QNM solutions. Effectively, $c_{n{\rm L}}$ is the contribution to the amplitude of the $n-$th QNM.

The solution \ref{psi} is a linear combination of different overtones. Keeping only the dominant $n=0$ mode gives us
\begin{eqnarray}
    \psi_0 = c_{0L}e^{Ts_0}\phi_{-}(x,s_0)\label{psi0}
\end{eqnarray}
as the solution to the first-order Zerilli equation.
$c_{0L}$ is the amplitude of the lowest (most dominant) QNM.
\section{Review of Second-order Solutions}\label{sec5}
Repeating the process with the second-order Zerilli equations, using the same Green's function, gives us the solutions in the notation of \cite{Perrone_2024} :
\begin{eqnarray}
\chi(x,T) &=& \chi^I(x,T) + \chi^S(x,T)\\
&=&\chi^I(x,T) + \chi^S(x,T)\Big|_{W \text{ poles}} + \chi^S(x,T)\Big|_{S \text{ poles}}.
\label{fullchi}
\end{eqnarray}
The three contributions come as we pick up the poles during the inverse Laplace transform of $\chi(x,s)$, with the source having two contributions: one is from the initial conditions at second order, and is similar to the previous section. The other corresponds to the source terms which are quadratic in the first order perturbation. The first term in \ref{fullchi}, $\chi^I(x,T)$, corresponds to the contribution  from the initial condition terms part of the source, and involves residues at the zeroes of the Wronskian while computing the inverse Laplace transform. The second term $\chi^S(x,T)\Big|_{W \text{ poles}}$, arises from the source, which depends on the first-order perturbation, and upon considering the residues at the zeros of the Wronskian. Finally the last term $\chi^S(x,T)\Big|_{S \text{ poles}}$, has the residues of the poles in the (Laplace transformed) source term. These come from the $s-2s_0$ term in the denominator of the source term after taking a Laplace transform, since the time dependence of the source term is $e^{2 s_0 T}$. In the Schwarzschild spacetime, if $s_0$ is a linear QNM frequency, $2s_0$ is not another linear QNM frequency, so we have a simple pole at $s_0$. The Laplace transformed source has the structure
\begin{equation}
S(x,s) = \frac{1}{s-2s_0} c_{0{\rm L}}^2 \; h(x),
\end{equation}
where $h(x)$ depends on the first-order perturbation.
All these terms have been discussed in \cite{Perrone_2024}. In their notation, the term corresponding to the initial conditions is
\begin{eqnarray}
\chi^I(x,T) = \sum_{n} \;c_{n{\rm NL}}\; \phi_-(x,s_n) \;e^{Ts_n} ,\label{chiI}
\end{eqnarray}
with
\begin{eqnarray}
    c_{n{\rm NL}}=W'(s_n)^{-1}\; \int_{-\infty}^{+\infty} \; {\rm d}x'\; \phi_-(x', s_n) \left( -s_n\,f_2(x') - g_2(x') \right).\label{cnNL}
\end{eqnarray}
Here, $s_n$ are the zeroes of the Wronskian.
The subscript `NL' stands for nonlinear (\textit{second order} perturbation). $c_{n{\rm NL}}$ counts the contribution to the solution from the initial conditions, and the poles of the Green's function, for the second order perturbation.
Next,
\begin{eqnarray}
\chi^S(x,T)\Big|_{W \text{ poles}}=\sum_{n}c_{W}e^{Ts_n}\phi_{-}(x,s_n),\label{chiw}
\end{eqnarray}
with
\begin{align}
c_W &= \frac{1}{W'(s_n)} \; \int_{-\infty}^{+\infty} \; {\rm d}x'\;  \phi_-(x', s_n) S(x',s_n) =\nonumber \\
&=\frac{c_{0{\rm L}}^2}{(s_n-2s_0) W'(s_n)} \int_{-\infty}^{+\infty} \; {\rm d}x'\;  \phi_-(x', s_n) \; h(x').\label{cw}
\end{align}
Once again, picking only the most dominant $n=0$ mode, gives us
\begin{equation}
    \chi^S(x,T)\Big|_{W \text{ poles}} = c_W \;e^{Ts_0}\; \phi_-(x,s_0), \label{chiWpole}
\end{equation}
with
\begin{align}
c_W &= \frac{c_{0{\rm L}}^2}{(-s_0) W'(s_0)} \int_{-\infty}^{+\infty} \; {\rm d}x'\;  \phi_-(x', s_0) \; h(x').
\end{align}
Thus, $c_W$ is the contribution to the second order perturbation from the source terms and the residues at the poles corresponding to the first zero of the Wronskian (neglecting contributions from higher order zeroes since they are more damped). $c_W$ contributes to the amplitude of the lowest linear QNM corresponding to the \textit{second order potential}.

Finally,
\begin{eqnarray}
    \chi^S(x,T)\Big|_{S \text{ poles}} = c_{0{\rm L}}^2 e^{2s_0 T} \; \int_{-\infty}^{+\infty} \; {\rm d}x'\;  \frac{\phi_+(x_>,2s_0)\, \phi_-(x_<, 2s_0)}{W(2s_0)} \; h(x').
\label{Spole}
\end{eqnarray}
This is the contribution we are after, since it is what contributes to the QQNM amplitude.
In this above analysis, we have assumed $s=2s_0$ is not a pole of the Wronskian. This is what we expect for QNMs in the Schwarzschild spacetime. This is distinct from \cite{Perrone_2024}, where their matching procedure amounts to assuming $2s_0$ is a linear QNM.

To compute the amplitude of the QQNM at frequency $2s_0$, we have to first evaluate \ref{Spole}. The contribution from the $W$ poles, \ref{chiWpole}, only renormalizes the linear QNM amplitude with frequency $s_0$. Of course, that also needs to be evaluated separately in order to compute the nonlinearity ratio.
Now, let us examine the source function $h(x)$ in the previous equation. It depends quadratically on the first order QNM, which naively is unbounded at the horizon, and also as $r \to \infty$. As has been emphasized by Szpak \cite{szpak2004quasinormalmodeexpansionexact}, and Okuzumi, Ioka and Sakagami\cite{Okuzumi:2008}, the actual QNMs have to be spatially truncated due to considerations of causality of the full time dependent problem. However, in these papers, the precise support of the spatially truncated QNMs are not identified, since that depends crucially on the form of the potential and is complicated to compute. This will be important to us in what follows.
Let us for now, assume the source is localized in a region $x_L < x' < x_R$. We are interested in finding the non-linearities in two regions: near the horizon and near infinity. Thus, we need the amplitude of the QQNM as $x \to -\infty$ and as $x \to \infty$.\vspace{2mm}\\
$\bold{CASE\quad1:}\quad x \rightarrow \infty$\vspace{2mm}\\
As in this case, we are specifically interested in the near-infinity region where $ x \rightarrow \infty$, we can safely say that we are always in the case where $x>x'$. Thus,
\begin{eqnarray}
 \chi^S(x \rightarrow \infty,T)\Big|_{S \text{ poles}} &=&c_{0{\rm L}}^2 e^{2s_0 T} \; \int_{-\infty}^{+\infty} \; {\rm d}x'\;  \frac{\phi_+(x_>,2s_0)\, \phi_-(x_<, 2s_0)}{W(2s_0)} \; h(x'), \\
 &=& c_{0{\rm L}}^2 e^{2s_0 T} \; \phi_+(x,2s_0) \; \int_{-\infty}^{+\infty} \; {\rm d}x'\;  \frac{\phi_-(x', 2s_0)}{W(2s_0)}\; h(x'),\label{csi}\\
 &=&  c_{Si} \;e^{2s_0 T} \; \phi_+(x,2s_0).\label{chiS}
\end{eqnarray}
where
\begin{equation}
    c_{Si}=  \frac{c_{0{\rm L}}^2}{W(2s_0)} \; \int_{-\infty}^{+\infty} \; {\rm d}x'\;  \phi_-(x', 2s_0)\; h(x').
\label{chiinfinity}
\end{equation}
Thus, $c_{Si}$ is the amplitude of the lowest quadratic quasinormal mode.

\vspace{3mm}

$\bold{CASE\quad2:}\quad x \rightarrow -\infty$\vspace{2mm}\\
Here, we are interested in the near-horizon region where $ x \rightarrow -\infty$, and thus we can assume $x<x'$. Similar to the earlier case,
\begin{eqnarray}
 \chi^S(x \rightarrow -\infty,T)\Big|_{S \text{ poles}} &=&c_{0{\rm L}}^2 e^{2s_0 T} \; \int_{-\infty}^{+\infty} \; {\rm d}x'\;  \frac{\phi_+(x_>,2s_0)\, \phi_-(x_<, 2s_0)}{W(2s_0)} \; h(x'),\\
 &=& c_{0{\rm L}}^2 e^{2s_0 T} \; \phi_-(x,2s_0) \; \int_{-\infty}^{+\infty} \; {\rm d}x'\;  \frac{\phi_+(x', 2s_0)}{W(2s_0)}\; h(x'), \label{csh}\\
 &=&  c_{Sh} \;e^{2s_0 T} \; \phi_-(x,2s_0),
 \end{eqnarray}
 where
 \begin{equation}
    c_{Sh}=  \frac{c_{0{\rm L}}^2}{W(2s_0)} \; \int_{-\infty}^{+\infty} \; {\rm d}x'\;  \phi_+(x', 2s_0)\; h(x').\label{chihorizon}
\end{equation}
$c_{Sh}$ is the amplitude of the lowest QQNM solution at the horizon.
Thus, the problem of finding the nonlinearities is reduced to finding $\phi_{\pm}(x,s)$, and then evaluating the integrals for $c_{Si}$ and $c_{Sh}$. We will do this in the WKB approximation.
\section{The WKB Approximation of Schutz and Will}\label{sec6}
Schutz and Will \cite{1985ApJ...291L..33S} evaluated QNMs in an analytical approximation scheme using WKB methods. We will need to summarize this, as we will later have to use such methods for computing second order perturbations.

Now, we know $\phi_-(x,s)$ and $\phi_+(x,s)$ are the homogeneous solutions of the Zerilli equation and thus follow
\begin{equation}
   \left[\partial_x^2  - s^2 - V(x) \right]\; \phi_{\pm}=0.
\end{equation}
Quasinormal modes are the solutions for which the reflection and transmission of the wave have approximately the same amplitude, and thus we can expect to find these solutions if we consider the case where the energy of the wave is comparable to the peak of the potential. That is, we require around the maximum $x_0$,
\begin{eqnarray}
    s^2 + V(x_0) \simeq 0.
\end{eqnarray}
Note that this holds only in the linear QNM case. Let us now define
\begin{equation}
    Q(x,s) = - s^2 - V(x),
\end{equation}
to rewrite our equation in the form in which we traditionally do the WKB approximation;
\begin{equation}
   \partial_x^2   \phi = -Q(x,s) \phi.
\end{equation}
Now, near the maximum, we can expand $V(x)$ around the maximum until the second order,
\begin{equation}
    Q(x,s) \sim -s^2 - V(x_0) - \frac{V''(x_0)}{2}(x-x_0)^2.
\end{equation}
As $V(x)$ has a maximum at $x_0$, we have,
\begin{eqnarray}
    V''(x_0) < 0,
\end{eqnarray}
and we further require that
\begin{equation}
    s^2 + V(x_0) > 0.
\end{equation}

\begin{figure}
    \centering
    \includegraphics[width=0.60\linewidth]{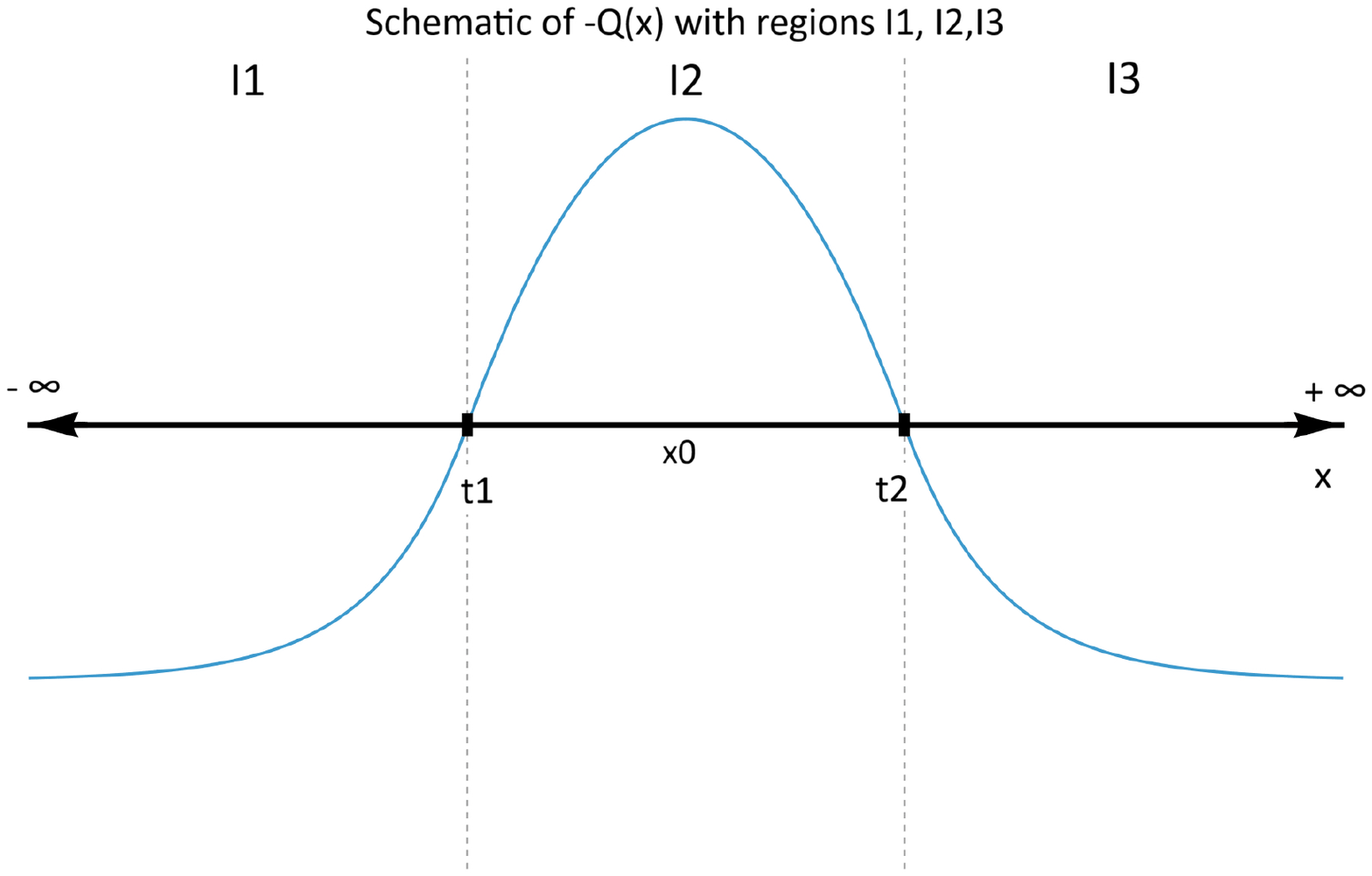}
    \caption{Diagram of $-Q(x)$}
    \label{regions}
\end{figure}

This is because we shall consider energies just below the maximum. Now, because of this, we can consider two turning points $t_1$ and $t_2$, such that
\begin{equation}
    s^2 + V(x)=0 \qquad \text{for     } x=t_1,t_2 .
\end{equation}
We can now identify the three different WKB regions:
 $I_1 = (-\infty, t_1)$ before the first turning point, $I_2 = (t_1, t_2)$ between the turning points, and $I_3 = (t_2,\infty)$ after the second turning point. The plot of $-Q(x)$ and the three regions are depicted in Figure \ref{regions}. We can now write the WKB solutions in these regions as given in \cite{1985ApJ...291L..33S}. Note that we are here using $e^{i\omega T}$ time dependence and hence $s =i\omega$\vspace{2mm}\\
 $\bullet\bold{\quad Region \quad I_1 = (-\infty, t_1)}$:
Here we will have the normal WKB solution before the first turning point, that is,
\begin{equation}
        \phi^{(1)}(x,s) = \frac{1}{Q^{1/4}(x,s)}\exp{\left(-i\int_{x}^{t_1} \sqrt{Q(x',s)}\;{\rm d}x'\right)}.\label{I1}
    \end{equation}
As $x \rightarrow -\infty$, $Q(x,s) = -s^2-V(x) \simeq -s^2$, as $V(x)\rightarrow 0$ here. Thus, we can write the limit of the solution in this region as
\begin{eqnarray}
    \phi^{(1)}(x,s) &=& \frac{1}{Q^{1/4}(x,s)}\exp{\left(-i\int_{x}^{t_1} \sqrt{Q(x',s)}\;{\rm d}x'\right)},\nonumber\\
    &\sim &\frac{1}{(\omega^2)^{1/4}}\exp{\left(-i\int_{x}^{t_1} \sqrt{\omega^2}\;{\rm d}x'\right)},\nonumber\\
    &=&\frac{1}{(\omega^2)^{1/4}}\exp{\left(-i\omega(t_1-x)\right)}, \nonumber\\
    &\sim&\frac{1}{(\omega^2)^{1/4}}e^{sx}. \label{I1 horizon}
\end{eqnarray}
Thus, as $x \rightarrow -\infty$, $\phi^{(1)}(x,s) \rightarrow \frac{1}{(\omega^2)^{1/4}}e^{sx}$.\vspace{2mm}\\
$\bullet\bold{\quad Region \quad I_3 = (t_2,\infty)}$:
Similar to before, here we will have the normal WKB solution after the second turning point,
    \begin{equation}
        \phi^{(3)}(x,s) = \frac{1}{Q^{1/4}(x,s)}\exp{\left(-i\int_{t_2}^{x} \sqrt{Q(x',s)}\;{\rm d}x'\right)}.\label{I3}
    \end{equation}
Again, as $x \rightarrow \infty$, $Q(x,s) = -s^2-V(x) \simeq -s^2$, as $V(x)\rightarrow 0$ here. Thus we can write the limit of the solution in this region as
\begin{eqnarray}
    \phi^{(3)}(x,s) &=& \frac{1}{Q^{1/4}(x,s)}\exp{\left(-i\int_{t_2}^{x} \sqrt{Q(x',s)}\;{\rm d}x'\right)}, \nonumber\\
    &\sim&\frac{1}{(\omega^2)^{1/4}}\exp{\left(-i\int_{t_2}^{x} \sqrt{\omega^2}\;{\rm d}x'\right)}, \nonumber\\
    &=&\frac{1}{(\omega^2)^{1/4}}\exp{\left(-i\omega(x-t_2)\right)}, \nonumber\\
    &\sim &\frac{1}{(\omega^2)^{1/4}}e^{-sx}.\label{phifar}
\end{eqnarray}
Thus, as $x \rightarrow \infty$, $\phi^{(3)}(x,s) \rightarrow \frac{1}{(-1)^{\frac{1}{4}}\sqrt{s}}e^{-sx}$.\vspace{2mm}\\
$\bullet\bold{\quad Region \quad I_2 = (t_1,t_2)}$:
Here we will solve the complete equation, with the potential expanded around the maximum  $x_0$, up to second order.
\begin{equation}
\left(\partial_x^2 - s^2 - V(x_0) - \frac 12 V''(x_0)(x-x_0)^2\right)\; \phi^{(2)} =0.\label{Z2WKB}
\end{equation}
Let us now introduce the substitutions,
\begin{eqnarray}
    k=-\frac 12 V''(x_0) = \frac 12 Q''(x_0),
\end{eqnarray}
\begin{eqnarray}
   t(x)=(4k)^{1/4} e^{i\pi/4} (x-x_0),
\end{eqnarray}
and
\begin{eqnarray}
    \nu = -i\frac{Q(x_0)}{\sqrt{2Q''(x_0)}}-\frac{1}{2} = i \frac{(s^2 + V(x_0))}{\sqrt{-2V''(x_0)}}-\frac{1}{2}. \label{nu}
\end{eqnarray}
Substituting this into \eqref{Z2WKB}, we get
\begin{equation}
\left(\partial_t^2 + \nu + \frac{1}{2} - \frac{1}{4}t^2\right)\; \phi^{(2)}(t) =0.
\end{equation}
This has solutions in terms of the parabolic cylinder functions,
\begin{equation}
\phi^{(2)}(t) = A D_{\nu}(t) + B D_{-1-\nu}(it).\label{I2}
\end{equation}
Now that we have the general solutions in the three regions, let us match the solutions. Here we shall use the property that the expansions of the parabolic cylinder functions are different as $x \rightarrow \infty$ and $x \rightarrow -\infty$.\vspace{2mm}\\
As $x \rightarrow \infty$,
\begin{eqnarray}
    \lim_{x\to +\infty}\phi^{(2)}(x) &=&   B e^{-3i\pi (\nu+1)/4} (4k)^{-(\nu+1)/4} (x-x_0)^{-(\nu+1)}e^{i \frac{\sqrt{k}}{2}(x-x_0)^2}, \nonumber\\
    &+&\left[ A + B \frac{\sqrt{2\pi} e^{-i \nu \pi /2}}{\Gamma(\nu + 1)}\right]e^{i \pi \nu /4} (4k)^{\nu/4} (x-x_0)^{\nu}e^{-i \frac{\sqrt{k}}{2}(x-x_0)^2}.\label{I2far}
\end{eqnarray}
As  $x \rightarrow -\infty$,
\begin{eqnarray}
    \lim_{x\to -\infty}\phi^{(2)}(x) &=&   A e^{-3i\pi \nu/4} (4k)^{\nu/4} (x_0-x)^{\nu}e^{-i \frac{\sqrt{k}}{2}(x-x_0)^2}, \nonumber \\
    &+&\left[ B -i A \frac{\sqrt{2\pi} e^{-i \nu \pi /2}}{\Gamma(-\nu )}\right]e^{i \pi (\nu+1) /4} (4k)^{-(\nu+1)/4} (x_0-x)^{-(\nu+1)}e^{i \frac{\sqrt{k}}{2}(x-x_0)^2}. \label{I2near}
\end{eqnarray}
This asymptotic expansion can be found in \cite{1985ApJ...291L..33S}. We have to match these to solutions in regions I1 and I3. While matching region I1 and I2, we match the far limit of the solution in I1 to the near limit of the solution in I2 at the first matching point $x_1$, similarly while matching region I2 and I3, we match the far limit of the solution in I2 to the the near limit of the solution in I3 at the second matching point $x_2$. In \cite{1985ApJ...291L..33S}, the matching points are just taken to be the exact turning points of the full potential. While computing the second-order perturbation, we will have to consider general matching points obeying certain inequalities. Henceforth, we will generically denote points at which we match asymptotic expansions as $x_1$ and $x_2$.

As $x \rightarrow -\infty$, we fall in the I1 region. Here the solution is
\begin{equation}
    \phi^{(1)}(x,s) = \frac{1}{Q^{1/4}(x,s)}\exp{\left(-i\int_{x}^{x_1} \sqrt{Q(x',s)}\;{\rm d}x'\right)}.
\end{equation}
To match this with the near limit of the I2 solution \eqref{I2near}, we thus need the second term to vanish. This is because, in this limit,
\begin{eqnarray}
    &&Q(x',s) = -s^2 - V(x_0) + k(x_0-x)^2 \simeq k(x_0-x)^2 ,\\
    &&\implies \sqrt{ Q(x',s)} \simeq \sqrt{k}(x_0-x),\\
    &&\implies \int_{x}^{x_1} \sqrt{Q(x',s)}\;{\rm d}x' \simeq \int_{x}^{x_1} \sqrt{k}(x_0-x)\;{\rm d}x' = \frac{\sqrt{k}}{2}[(x_0-x)^2 - (x_0-x_1)^2] \simeq \frac{\sqrt{k}}{2}(x_0-x)^2 .
\end{eqnarray}
Thus,
\begin{eqnarray}
    \phi^{(1)}(x,s) = \frac{1}{Q^{1/4}(x,s)}\exp{\left(-i\int_{x}^{x_1} \sqrt{Q(x',s)}\;{\rm d}x'\right)} \rightarrow e^{-i\sqrt{k}(x_0-x)^2}.
\end{eqnarray}

For this to occur, we need to impose the condition
\begin{eqnarray}
    B = i A \frac{\sqrt{2\pi} e^{-i \nu \pi /2}}{\Gamma(-\nu )}.
\label{nearcondn}
\end{eqnarray}
Now, we can get the correct matching coefficients for $\phi_-$ at $x_1$, by equating the near limit of the I2 solution \eqref{I2near} to the I1 solution \eqref{I1} at $x_1$.
\begin{eqnarray}
    &&A_- e^{-3i\pi \nu/4} (4k)^{\nu/4} (x_0-x_1)^{\nu}e^{-i \frac{\sqrt{k}}{2}(x_1-x_0)^2} = \frac{1}{Q^{1/4}(x_1,s)},\\
    &\implies& A_- = \frac{ e^{3i\pi \nu/4} (4k)^{-\nu/4} (x_0-x_1)^{-\nu}e^{i \frac{\sqrt{k}}{2}(x_1-x_0)^2}}{Q^{1/4}(x_1,s)}. \label{AMQNM}
\end{eqnarray}
Now, matching the far limit of the I2 solution \eqref{I2far} to the I3 solution \eqref{I3}, to have the correct exponentials, we have to impose
\begin{eqnarray}
    B=0 \implies \frac{1}{\Gamma(-\nu)}=0 \implies \nu \in \mathbb{N}_0 ,
\end{eqnarray}
where the implication comes from \ref{nearcondn}. We can use the formula \ref{nu} to get the $s_n$ for which this is true. Thus, for $s=s_n$, the solution is ingoing at the horizon \textit{and } simultaneously outgoing at infinity. This is thus the QNM solution.
Now, equating the far limit of I2 solution \eqref{I2far} to the I3 solution \eqref{I3} at $x_2$,
\begin{eqnarray}
     &&A_-e^{i \pi \nu /4} (4k)^{\nu/4} (x_2-x_0)^{\nu}e^{-i \frac{\sqrt{k}}{2}(x_2-x_0)^2} = \frac{C}{Q^{1/4}(x_2,s)},\\
     &\implies& C = A_-Q^{1/4}(x_2,s)e^{i \pi \nu /4} (4k)^{\nu/4} (x_2-x_0)^{\nu}e^{-i \frac{\sqrt{k}}{2}(x_2-x_0)^2}.
\end{eqnarray}
Thus, the QNM solution is
\begin{equation}
    \phi_-(x,s_n) =
    \begin{cases}
        \frac{1}{Q^{1/4}(x,s_n)}\exp{(-i\int_{x}^{x_1} \sqrt{Q(x',s_n)}\;{\rm d}x')},\quad &x<x_1\\
        A_- D_n(t(x)), \quad &x_1<x<x_2\\
        \frac{C}{Q^{1/4}(x,s)}\exp\left({-i\int_{x_2}^{x} \sqrt{Q(x',s)}\;{\rm d}x'}\right),\quad &x_2<x .\label{phi-qnm}
    \end{cases}
\end{equation}
We can do the same for $\phi_+$, which is expected to be linearly dependent to $\phi_-$. We will spell out the steps, since we will have to do this procedure again at second order.
Matching the far limit of the I2 solution \eqref{I2far} to the I3 solution \eqref{I3} gives us
\begin{eqnarray}
    B=0,
\end{eqnarray}
and equating them at $x_2$ gives us
\begin{eqnarray}
    &&A_+ e^{i \pi \nu /4} (4k)^{\nu/4} (x_2-x_0)^{\nu}e^{-i \frac{\sqrt{k}}{2}(x_2-x_0)^2} =  \frac{1}{Q^{1/4}(x_2,s)} \nonumber \\
    &\implies& A_+ = \frac{e^{-i \pi \nu /4} (4k)^{-\nu/4} (x_2-x_0)^{-\nu}e^{i \frac{\sqrt{k}}{2}(x_2-x_0)^2}}{Q^{1/4}(x_2,s)}. \label{aplusQNM}
\end{eqnarray}
Now, equating the near limit of I2 solution \eqref{I2near} to the I1 solution \eqref{I1} at $x_1$ for $\phi_+$ gives us
\begin{eqnarray}
    &&A_+ e^{-3i\pi \nu/4} (4k)^{\nu/4} (x_0-x_1)^{\nu}e^{-i \frac{\sqrt{k}}{2}(x_1-x_0)^2} = \frac{D}{Q^{1/4}(x_1,s)},\\
    &\implies& D = A_+Q^{1/4}(x_1,s) e^{-3i\pi \nu/4} (4k)^{\nu/4} (x_0-x_1)^{\nu}e^{-i \frac{\sqrt{k}}{2}(x_1-x_0)^2}.
\end{eqnarray}
Thus, we have
\begin{equation}
    \phi_+(x,s_n) =
    \begin{cases}
        \frac{D}{Q^{1/4}(x,s_n)}\exp{(-i\int_{x}^{x_1} \sqrt{Q(x',s_n)}\;{\rm d}x')},\quad &x<x_1\\
        A_+ D_n(t(x)), \quad &x_1<x<x_2\\
        \frac{1}{Q^{1/4}(x,s)}\exp\left({-i\int_{x_2}^{x} \sqrt{Q(x',s)}\;{\rm d}x'}\right).\quad &x_2<x
    \end{cases}
\end{equation}
We see that $\phi_+$ and $\phi_-$ are linearly dependent for $s = s_n$, and hence the Wronskian of the two solutions $W(s_n) = 0$. The exact values of the $s_n$ in this WKB approximation scheme can be found in \cite{1985ApJ...291L..33S} and these values closely match QNM frequencies computed numerically.

Our goal is to compute \ref{chiinfinity} and \ref{chihorizon}, and compute the nonlinearities from them. For this, we need to compute solutions for $\phi_+(x,s)$ and $\phi_-(x,s)$ when $s \neq s_n$ --- we will mostly need $\phi_{\pm}(x,s)$ for $s = 2s_0$. We proceed to do this in the next section.
\section{General solution for $\phi_+(x,s)$ for $s \neq s_n$ }\label{sec7}
We now want to construct an approximate solution to the homogeneous equation (second order Zerilli equation with source set to zero) which obeys the QNM boundary condition at infinity. We would like to take $s \neq s_n$ (i.e., $s$ is not a linear QNM frequency of the final $l$ value), so it is a linear combination of both ingoing and outgoing solutions at the horizon. In order to do this, we will match solutions in I2 and I3 at an intermediate point $x_2$, and then study this matched solution in I1 by matching it to the linear combination of linearly independent solutions in I1 at a point $x_1$. $x_1$, $x_2$ are chosen such that for $s=2s_0$, they obey the relations
\begin{eqnarray}
Q(x_1,s) = -s^2-V(x_0)+k(x_0-x_1)^2 \simeq k(x_0-x_1)^2 ,\\
Q(x_2,s) = -s^2-V(x_0)+k(x_2-x_0)^2 \simeq k(x_2-x_0)^2 .\label{Q approx}
\end{eqnarray}
Further, we must choose $x_1$, $x_2$ such that at these points, the cubic term in the Taylor expansion of the potential about $x_0$ is smaller than the quadratic term. We need these conditions to hold, in order to be able to match solutions across regions when $s \neq s_n$.
These conditions imply an allowed range of matching points.\\
Defining
$$t=x_0-x_1 = x_2-x_0,$$
we get an upper bound on $t$, as the quadratic term in the WKB approximation must be greater than the cubic term, that is,
\begin{eqnarray}
    \left|\frac{V''(x_0)}{2}t^2\right| > \left|\frac{V'''(x_0)}{6}t^3\right|,
\end{eqnarray}
and also a lower bound on $t$ to apply the approximation \eqref{Q approx}, that is,
\begin{eqnarray}
    \left|-s^2 - V(x_0)\right|< \left|\frac{V''(x_0)}{2}t^2\right|.
\end{eqnarray}
This gives us an allowed range for $x_1,x_2$, the matching points at linear order, and $y_1,y_2$, the matching points at nonlinear order.  Later, we will discuss the sensitivity of the nonlinearity ratio to the choice of matching points.\\
Now, we already know that
\begin{equation}
    \phi_+(x,s) =
    \begin{cases}
          \frac{\tilde{A}}{{Q(x,s)}^{\frac{1}{4}}}\exp{\left(-i\int_{x}^{x_1} \sqrt{Q(x',s)}\;{\rm d}x'\right)} + \frac{\tilde{B}}{{Q(x,s)}^{\frac{1}{4}}}\exp{\left(i\int_{x}^{x_1} \sqrt{Q(x',s)}\;{\rm d}x'\right)}&x<x_1,\\
        A_+ D_{\nu}(t(x))+B_+D_{-1-\nu}(it), \quad &x_1<x<x_2,\\
         \frac{1}{Q^{1/4}(x,s)}\exp{(-i\int_{x_2}^{x} \sqrt{Q(x',s)}\;{\rm d}x')}\quad &x_2<x .
    \end{cases}
    \label{patchsoln}
\end{equation}
Taking the far limit of the I2 solution given by \eqref{I2far}, we have, \vspace{2mm}\\
as $x \rightarrow \infty$:
\begin{eqnarray}
    \lim_{x\to +\infty}\phi^{(2)}_+(x) &=&   B_+ e^{-3i\pi (\nu+1)/4} (4k)^{-(\nu+1)/4} (x-x_0)^{-(\nu+1)}e^{i \frac{\sqrt{k}}{2}(x-x_0)^2} \nonumber \\
    &+&\left[ A_+ + B_+ \frac{\sqrt{2\pi} e^{-i \nu \pi /2}}{\Gamma(\nu + 1)}\right]e^{i \pi \nu /4} (4k)^{\nu/4} (x-x_0)^{\nu}e^{-i \frac{\sqrt{k}}{2}(x-x_0)^2}.
\end{eqnarray}
To ensure purely outgoing behaviour, we thus put $B_+=0$. Now, equating the far limit of I2 solution \eqref{I2far} to the I3 solution \eqref{I3} at $x_2$,
\begin{eqnarray}
    &&A_+ e^{i \pi \nu /4} (4k)^{\nu/4} (x_2-x_0)^{\nu}e^{-i \frac{\sqrt{k}}{2}(x_2-x_0)^2} = \frac{1}{Q^{1/4}(x_2,s)},\\
    &\implies& A_+ = \frac{e^{-i \pi \nu /4} (4k)^{-\nu/4} (x_2-x_0)^{-\nu}e^{i \frac{\sqrt{k}}{2}(x_2-x_0)^2}}{Q^{1/4}(x_2,s)}. \label{AP}
\end{eqnarray}
Now, taking the near limit of the I2 solution \eqref{I2near}, we have, as $x \rightarrow -\infty$;
\begin{eqnarray}
    \lim_{x\to -\infty}\phi^{(2)}_+(x) &=&   A_+ e^{-3i\pi \nu/4} (4k)^{\nu/4} (x_0-x)^{\nu}e^{-i \frac{\sqrt{k}}{2}(x-x_0)^2} \nonumber \\
    &+&\left[ -i A_+ \frac{\sqrt{2\pi} e^{-i \nu \pi /2}}{\Gamma(-\nu )}\right]e^{i \pi (\nu+1) /4} (4k)^{-(\nu+1)/4} (x_0-x)^{-(\nu+1)}e^{i \frac{\sqrt{k}}{2}(x-x_0)^2}.
\end{eqnarray}
We see, that as expected, it has both ingoing and outgoing parts as in \ref{patchsoln}, since it is not itself a linear QNM ($s \neq s_n$).
Equating this to the I1 solution \eqref{I1} at $x_1$, we have;
\begin{eqnarray}
    &&\frac{\tilde{A}}{Q^{1/4}(x_1,s)} = A_+ e^{-3i\pi \nu/4} (4k)^{\nu/4} (x_0-x_1)^{\nu}e^{-i \frac{\sqrt{k}}{2}(x_1-x_0)^2}, \\
    &\implies& \tilde{A} = Q^{1/4}(x_1,s)A_+ e^{-3i\pi \nu/4} (4k)^{\nu/4} (x_0-x_1)^{\nu}e^{-i \frac{\sqrt{k}}{2}(x_1-x_0)^2}. \label{AT}
\end{eqnarray}
Further,
\begin{eqnarray}
    &&\frac{\tilde{B}}{Q^{1/4}(x_1,s)} =\left[ -i A_+ \frac{\sqrt{2\pi} e^{-i \nu \pi /2}}{\Gamma(-\nu )}\right]e^{i \pi (\nu+1) /4} (4k)^{-(\nu+1)/4} (x_0-x_1)^{-(\nu+1)}e^{i \frac{\sqrt{k}}{2}(x_1-x_0)^2},\\
    &\implies&\tilde{B} = Q^{1/4}(x_1,s)\left[ -i A_+ \frac{\sqrt{2\pi} e^{-i \nu \pi /2}}{\Gamma(-\nu )}\right]e^{i \pi (\nu+1) /4} (4k)^{-(\nu+1)/4} (x_0-x_1)^{-(\nu+1)}e^{i \frac{\sqrt{k}}{2}(x_1-x_0)^2}. \label{BT}
\end{eqnarray}
This gives us
\begin{equation}
    \phi_+(x,s) =
    \begin{cases}
          \frac{\tilde{A}}{{Q(x,s)}^{\frac{1}{4}}}\exp{\left(-i\int_{x}^{x_1} \sqrt{Q(x',s)}\;{\rm d}x'\right)} + \frac{\tilde{B}}{{Q(x,s)}^{\frac{1}{4}}}\exp{\left(i\int_{x}^{x_1} \sqrt{Q(x',s)}\;{\rm d}x'\right)}&x<x_1 ,\\
        A_+ D_{\nu}(t(x)), \quad &x_1<x<x_2, \\
         \frac{1}{Q^{1/4}(x,s)}\exp{(-i\int_{x_2}^{x} \sqrt{Q(x',s)}\;{\rm d}x')},\quad &x_2<x ,\label{phi+}
    \end{cases}
\end{equation}
with $\tilde A$ and $\tilde B$ determined by \ref{AT} and \ref{BT}, and $A_+$ determined by \ref{AP}.
\section{General solution for $\phi_-(x,s)$ for $s \neq s_n$ }\label{sec8}
We now do the same with the solution that is purely ingoing at the horizon, and is a linear combination of the ingoing and outgoing solutions at infinity. This is distinct from the matching procedure in \cite{Perrone_2024}, where they match such that $s$ is one of the linear QNMs and such that $\nu = n$. We would like to set $s = 2s_0$ while computing nonlinearities, but $2s_0$ is not a linear QNM in general. Therefore, we do not assume $s$ to be a linear QNM. We strictly have $s \neq s_n$.
We can write the solution as
\begin{equation}
    \phi_-(x,s) =
    \begin{cases}
        \frac{1}{Q^{1/4}(x,s)}\exp{(-i\int_{x}^{x_1} \sqrt{Q(x',s)}\;{\rm d}x')},\quad &x<x_1\\
        aD_{\nu}(t)+bD_{-1-\nu}(it), \quad &x_1<x<x_2\\
        \frac{\tilde{C}}{Q^{1/4}(x,s)}\exp\left({-i\int_{x_2}^{x} \sqrt{Q(x',s)}\;{\rm d}x'}\right)+\frac{\tilde{D}}{Q^{1/4}(x,s)}\exp\left({i\int_{x_2}^{x} \sqrt{Q(x',s)}\;{\rm d}x'}\right).\quad &x_2<x\label{phi-}
    \end{cases}
\end{equation}
Taking the near limit of the I2 solution \eqref{I2near}, we have
as  $x \rightarrow -\infty$:
\begin{eqnarray}
    \lim_{x\to -\infty}\phi^{(2)}_-(x) &=&   a e^{-3i\pi \nu/4} (4k)^{\nu/4} (x_0-x)^{\nu}e^{-i \frac{\sqrt{k}}{2}(x-x_0)^2} \nonumber \\
    &+&\left[ b -i a \frac{\sqrt{2\pi} e^{-i \nu \pi /2}}{\Gamma(-\nu )}\right]e^{i \pi (\nu+1) /4} (4k)^{-(\nu+1)/4} (x_0-x)^{-(\nu+1)}e^{i \frac{\sqrt{k}}{2}(x-x_0)^2}.
\end{eqnarray}
To ensure purely ingoing behaviour, we put
\begin{eqnarray}
    b = i a \frac{\sqrt{2\pi} e^{-i \nu \pi /2}}{\Gamma(-\nu )}\label{b}
\end{eqnarray}
Now equating the near limit of the I2 solution \eqref{I2near} to the I1 solution \eqref{I1} at $x_1$, we have,
\begin{eqnarray}
    &&a e^{-3i\pi \nu/4} (4k)^{\nu/4} (x_0-x_1)^{\nu}e^{-i \frac{\sqrt{k}}{2}(x_1-x_0)^2} = \frac{1}{Q^{1/4}(x_1,s)},\\
    &\implies& a = \frac{e^{3i\pi \nu/4} (4k)^{-\nu/4} (x_0-x_1)^{-\nu}e^{i \frac{\sqrt{k}}{2}(x_1-x_0)^2}}{Q^{1/4}(x_1,s)}.\label{asol}
\end{eqnarray}
Next, taking the far limit of the I2 solution \eqref{I2far}, we have as $x \rightarrow \infty$;
\begin{eqnarray}
    \lim_{x\to +\infty}\phi^{(2)}_-(x) &=&   b e^{-3i\pi (\nu+1)/4} (4k)^{-(\nu+1)/4} (x-x_0)^{-(\nu+1)}e^{i \frac{\sqrt{k}}{2}(x-x_0)^2} \nonumber \\
    &+&\left[ a + b \frac{\sqrt{2\pi} e^{-i \nu \pi /2}}{\Gamma(\nu + 1)}\right]e^{i \pi \nu /4} (4k)^{\nu/4} (x-x_0)^{\nu}e^{-i \frac{\sqrt{k}}{2}(x-x_0)^2}.
\end{eqnarray}
Equating this to the I3 solution \eqref{I3} at $x_2$, we have,
\begin{eqnarray}
    \left[ a + b \frac{\sqrt{2\pi} e^{-i \nu \pi /2}}{\Gamma(\nu + 1)}\right]e^{i \pi \nu /4} (4k)^{\nu/4} (x_2-x_0)^{\nu}e^{-i \frac{\sqrt{k}}{2}(x_2-x_0)^2} = \frac{\tilde{C}}{Q^{1/4}(x_2,s)}.
\end{eqnarray}
This gives us
\begin{eqnarray}
    \tilde{C} = \left[ a + b \frac{\sqrt{2\pi} e^{-i \nu \pi /2}}{\Gamma(\nu + 1)}\right]e^{i \pi \nu /4} (4k)^{\nu/4} (x_2-x_0)^{\nu}e^{-i \frac{\sqrt{k}}{2}(x_2-x_0)^2} Q^{\frac{1}{4}}(x_2,s).\label{CT}
\end{eqnarray}
We also have
\begin{eqnarray}
    b e^{-3i\pi (\nu+1)/4} (4k)^{-(\nu+1)/4} (x_2-x_0)^{-(\nu+1)}e^{i \frac{\sqrt{k}}{2}(x_2-x_0)^2} = \frac{\tilde{D}}{Q^{1/4}(x_2,s)}.
\end{eqnarray}
This gives us
\begin{eqnarray}
    \tilde{D} = \tilde{B}.
\end{eqnarray}
{\bf A cross-check:}\\
Using the $b$ that we have from \eqref{b}, let us now also calculate the Wronskian at $x_0$, as a cross-check of this matching procedure. Recall that for equations of the Zerilli type (i.e., Schrodinger type), the Wronskian should be independent of $x$. Using the solutions \eqref{phi+} and \eqref{phi-}, we have

\begin{eqnarray}
    W(x_0) &=& -ibA_+(4k)^{1/4}e^{i\pi/4}e^{-i\pi\nu/2} \nonumber\\
     &=& -i (4k)^{1/4}e^{i\pi/4}e^{-i\pi\nu/2} \left[i a \frac{\sqrt{2\pi} e^{-i \nu \pi /2}}{\Gamma(-\nu )}\right]A_+ \nonumber \\
     &=&-i (4k)^{1/4}e^{i\pi/4}e^{-i\pi\nu/2} \left[i \left[\frac{e^{3i\pi \nu/4} (4k)^{-\nu/4} (x_0-x_1)^{-\nu}e^{i \frac{\sqrt{k}}{2}(x_1-x_0)^2}}{Q^{1/4}(x_1,s)}\right] \frac{\sqrt{2\pi} e^{-i \nu \pi /2}}{\Gamma(-\nu )}\right]A_+ \nonumber \\
     &=&\frac{(4k)^{1/4}e^{i\pi/4}e^{-i\pi\nu/2}e^{3i\pi \nu/4} (4k)^{-\nu/4}  (x_0-x_1)^{-\nu}e^{i \frac{\sqrt{k}}{2}(x_1-x_0)^2}\sqrt{2\pi} e^{-i \nu \pi /2}}{\Gamma(-\nu)k^{\frac{1}{4}}(x_0-x_1)^\frac{1}{2}}A_+ \nonumber \\
     &=&\frac{(4)^{1/4}e^{i\pi/4}e^{i\pi \nu/4} (4k)^{-\nu/4}  (x_0-x_1)^{-\nu-\frac{1}{2}}e^{i \frac{\sqrt{k}}{2}(x_1-x_0)^2}\sqrt{2\pi} e^{-i \nu \pi /2}}{\Gamma(-\nu)}A_+ \nonumber \\
     &=&\frac{(4)^{1/4}(x_0-x_1)^{\frac{1}{2}}e^{i\pi \frac{\nu+1}{4}} (4k)^{\frac{1}{4}}(4k)^{-\frac{\nu+1}{4}}  (x_0-x_1)^{-\nu-1}e^{i \frac{\sqrt{k}}{2}(x_1-x_0)^2}\sqrt{2\pi} e^{-i \nu \pi /2}}{\Gamma(-\nu)}A_+ \nonumber \\
     &=&\frac{2(k)^{\frac{1}{4}}(x_0-x_1)^{\frac{1}{2}}e^{i\pi \frac{\nu+1}{4}} (4k)^{-\frac{\nu+1}{4}}  (x_0-x_1)^{-\nu-1}e^{i \frac{\sqrt{k}}{2}(x_1-x_0)^2}\sqrt{2\pi} e^{-i \nu \pi /2}}{\Gamma(-\nu)}A_+ \nonumber \\
     &=&\frac{2Q^{\frac{1}{4}}(x_1,s)e^{i\pi \frac{\nu+1}{4}} (4k)^{-\frac{\nu+1}{4}}  (x_0-x_1)^{-\nu-1}e^{i \frac{\sqrt{k}}{2}(x_1-x_0)^2}\sqrt{2\pi} e^{-i \nu \pi /2}}{\Gamma(-\nu)}A_+ .
\end{eqnarray}
Here, we use \eqref{Q approx} to write
\begin{eqnarray}
     Q^{\frac{1}{4}}(x_1,s) \simeq k^{\frac{1}{4}}(x_0-x_1)^{\frac{1}{2}}.
\end{eqnarray}

Therefore,
\begin{eqnarray}
     W(x_0) &=& 2 i\frac{-iQ^{\frac{1}{4}}(x_1,s)e^{i\pi \frac{\nu+1}{4}} (4k)^{-\frac{\nu+1}{4}}  (x_0-x_1)^{-\nu-1}e^{i \frac{\sqrt{k}}{2}(x_1-x_0)^2}\sqrt{2\pi} e^{-i \nu \pi /2}}{\Gamma(-\nu)}A_+\nonumber \\
     &=& 2 i Q^{\frac{1}{4}} \left[-iA_+\frac{\sqrt{2\pi}e^{-\frac{i\pi\nu}{2}}}{\Gamma(-\nu)}\right]e^{i\pi \frac{\nu+1}{4}} (4k)^{-\frac{\nu+1}{4}}(x_0-x_1)^{-\nu-1}e^{i \frac{\sqrt{k}}{2}(x_1-x_0)^2} \nonumber \\
     &=&2 i \tilde{B}\label{W}.
\end{eqnarray}
Now, let us calculate the Wronskian at the horizon.
\begin{eqnarray}
    \phi_-(x\rightarrow -\infty,s) \rightarrow \frac{e^{sx}}{(\omega^2)^{\frac{1}{4}}};
\end{eqnarray}
and
\begin{eqnarray}
    \phi_+(x \rightarrow \infty,s) \rightarrow \frac{\tilde{A}e^{sx}}{(\omega^2)^{\frac{1}{4}}} +\frac{\tilde{B}e^{-sx}}{(\omega^2)^{\frac{1}{4}}}.
\end{eqnarray}
Therefore we get the Wronskian to be $2i\tilde{B}$, which matches the Wronskian at $x_0$. Doing this calculation at infinity also gives us the same Wronskian.

\section{Calculating non-linearities for $l=2$}\label{sec9}
In this section we shall evaluate the non-linearities for the $(2,0)\times(2,0)\rightarrow(2,0)$ channel. This channel is excited by a head-on collision of two spinless black holes. We will attempt to do this in an analytical approximation, using the method of steepest descent. Unfortunately, the exact contour of steepest descent is hard to find. For a naive choice of contour as for example used in PBKR, we discuss why the approximation fails. However, as we see in the next section, for a different channel (and perhaps the most important channel for QQNMs), steepest descent works for this contour choice.
The reason we do the full computation in this section and then discuss its failure, is that this example is more involved than other channels due to the renormalization of the linear parent mode.

Further, we can go to dimensionless coordinates by scaling the coordinates by a factor of $1/M$. Writing the nonlinearity ratio in terms of dimensionless coordinates, we can check that the $M$ dependence is scaled out (this can be explicitly also checked in the source term \ref{Sriotto}. Henceforth, in computations, we will assume all quantities can be expressed in these dimensionless variables, or equivalently, set $M=1$.

Let us now write
\begin{eqnarray}
    h(x) = H(x)\phi_-^2(x,s_0), \end{eqnarray}
where, from \eqref{phi-qnm} we have
\begin{eqnarray}
    \phi_-(x,s_0)=A_-(s_0)D_0(t(x)) = A_-(s_0)e^{-\frac{i\sqrt{k}(x-x_0)^2}{2}}.
\end{eqnarray}
 The parabolic cylinder function can be written in terms of confluent hypergeometric functions as
\begin{eqnarray}
    D_{a}[z] = \frac{2^{\frac{a}{2}}\sqrt{\pi}}{\Gamma(\frac{1-a}{2})}e^{-\frac{z^2}{4}}\;{}_1F_1(\frac{-a}{2},\frac{1}{2},\frac{z^2}{2})-\frac{\sqrt{2 \pi}2^{\frac{a}{2}}}{\Gamma(-\frac{a}{2})}ze^{-\frac{z^2}{4}}\;{}_1F_1(\frac{1-a}{2},\frac{3}{2},\frac{z^2}{2}).
\end{eqnarray}
Now, we know from \eqref{csi} that
\begin{eqnarray}
    \chi^S(x \rightarrow \infty,T)|_{S \text{ pole}} = \frac{c_{0{\rm L}}^2 e^{2s_0 T}  \phi_+(x,2s_0)}{W(2s_0)} \; \int_{-\infty}^{+\infty} \; {\rm d}x'\;  \phi_-(x', 2s_0)\; h(x') ,
\end{eqnarray}
which gives us nonlinearity at infinity defined as the ratio of the non-linear amplitude divided by the square of the linear amplitudes. Further, as nonlinearity is defined as the ratio of the metric perturbations instead of the Zerilli function $\chi$ we get an extra $\frac{1}{2s_0}$ factor. This is because the relation between them at first order is:
\begin{eqnarray}
    \psi_1 = \psi,
\end{eqnarray}
and at second order,
\begin{eqnarray}
    \dot{\psi_2} = \chi^S \implies \psi_2 = \frac{\chi^S}{2s_0}.
\end{eqnarray}
Here, $\chi^S$ is short-hand for $\chi^S(x \rightarrow \infty,T)|_{S \text{ pole}}$, which is the amplitude of the QQNM at frequency $2s_0$.
This gives us the final nonlinearity ratio
\begin{eqnarray}
    NL(x \rightarrow \infty) = \left|\frac{\psi_2}{\psi_1^2}\right|= \left|\frac{1}{2s_0}\frac{\chi^S}{\psi_1^2} \right|.
\end{eqnarray}
However, as shown in \cite{Perrone_2024}, the denominator also gets renormalized by the contribution at second order of $\chi^I$ and $\chi^S\Big|_{W \text{ poles}}$, as given by \eqref{chiI} and \eqref{chiWpole} to the amplitude of the linear QNM at $s=s_0$. This is because two seed $(2,0)$ modes are giving us again a $(2,0)$ mode at second order. Thus the observed nonlinearity is
\begin{eqnarray}
    NL(x \rightarrow \infty) &=& \left|\frac{1}{2s_0}\frac{\chi}{\psi_1^2}\right|\;\frac{1}{\left|1+\frac{c_{0NL}}{c_{0L}}+\frac{c_w}{c_{0L}}\right|^2}, \nonumber \\
    &=& \left|\frac{1}{2s_0}\frac{c_{0{\rm L}}^2 e^{2s_0 T}  \phi_+(x\rightarrow\infty,2s_0)}{W(2s_0)c_{0L}^2 e^{2s_0 T}\phi_-^2(x\rightarrow \infty,s_0)}\int_{-\infty}^{+\infty} \; {\rm d}x'\;  \phi_-(x', 2s_0)\; h(x')\right|\;\frac{1}{\left|1+\frac{c_{0NL}}{c_{0L}}+\frac{c_w}{c_{0L}}\right|^2},\nonumber \\
    &=&\left|\frac{1}{2s_0}\sqrt{\frac{s_0}{2}}\frac{1}{W(2s_0)}\frac{1}{\left|1+\frac{c_{0NL}}{c_{0L}}+\frac{c_w}{c_{0L}}\right|^2}\int_{-\infty}^{+\infty} \; {\rm d}x'\;  \phi_-(x', 2s_0)\; h(x')\right| .\label{NLI}
\end{eqnarray}
Let us now look at the integral, using the solution for $\phi_-(x,s)$ from \eqref{phi-},
\begin{eqnarray}
    &&\int_{-\infty}^{+\infty} \; {\rm d}x'\;  \phi_-(x', 2s_0)\; h(x'), \nonumber \\
    &=&\int_{-\infty}^{+\infty} \; {\rm d}x'\;  [aD_{\nu}(t')+bD_{-1-\nu}(it')] A_-^2 D_0^2(t')H(x'), \nonumber \\
    &=&\int_{-\infty}^{+\infty} \; {\rm d}x'\;  [aD_{\nu}(t')+bD_{-1-\nu}(it')] A_-^2 D_0^2(t')H(x'),
\label{NLintegral}
\end{eqnarray}
where,
\begin{eqnarray}
    h(x) = H(x)\phi_-^2(x,s_0),
\end{eqnarray}
and
\begin{eqnarray}
    t' = t(x').
\end{eqnarray}
The approximation used for example in \cite{Perrone_2024} is as follows: the actual QNMs are truncated spatially from considerations of causality as already discussed. Therefore, we can approximate the integral in \ref{NLintegral} to be only over region I2, and we can replace $\phi_{-}(x', 2s_0)$ and $h(x')$ by their forms in this region.
Finally, applying the method of steepest descent about the point $x_0$ to evaluate this integral gives us,
\begin{eqnarray}
    &=& H(x_0)a A_-^2\frac{2^{\frac{\nu}{2}}\sqrt{\pi}}{\Gamma(\frac{1-\nu}{2})}\int_{-\infty}^{+\infty} \; {\rm d}x'e^{\frac{-3i\sqrt{k}}{2}(x'-x_0)^2} + H(x_0)bA_-^2\frac{2^{\frac{-1-\nu}{2}}\sqrt{\pi}}{\Gamma(1+\frac{\nu}{2})}\int_{-\infty}^{+\infty} \; {\rm d}x'e^{\frac{-i\sqrt{k}}{2}(x'-x_0)^2}, \nonumber \\
    &=& H(x_0)a A_-^2\frac{2^{\frac{\nu}{2}}\sqrt{\pi}}{\Gamma(\frac{1-\nu}{2})}\sqrt{\frac{2\pi}{3i\sqrt{k}}}+H(x_0)bA_-^2\frac{2^{\frac{-1-\nu}{2}}\sqrt{\pi}}{\Gamma(1+\frac{\nu}{2})}\sqrt{\frac{2\pi}{i\sqrt{k}}}.
    \label{imgaussian}
\end{eqnarray}
Further, we also have from \eqref{cw},
\begin{eqnarray}
    c_W &=& \frac{1}{W'(s_n)} \; \int_{-\infty}^{+\infty} \; {\rm d}x'\;  \phi_-(x', s_n) S(x',s_n), \nonumber \\
    &=&\frac{c_{0{\rm L}}^2}{(s_n-2s_0) W'(s_n)} \int_{-\infty}^{+\infty} \; {\rm d}x'\;  \phi_-(x', s_n) \; h(x'), \nonumber \\
    &=&\frac{c_{0{\rm L}}^2}{(-s_0) W'(s_0)} \int_{-\infty}^{+\infty} \; {\rm d}x'\;  \phi_-(x', s_n) \; h(x').
\end{eqnarray}
Solving this integral using the method of steepest descent gives us,
\begin{eqnarray}
     \int_{-\infty}^{+\infty} \; {\rm d}x'\;  \phi_-(x', s_n) \; h(x')
     &=& A_-^3(s_0)\int_{-\infty}^{+\infty} \; {\rm d}x'H(x')e^{\frac{-3i\sqrt{k}}{2}(x'-x_0)^2}, \nonumber \\
     &=&A_-^3(s_0)H(x_0)\sqrt{\frac{2\pi}{3i\sqrt{k}}}.
\end{eqnarray}
Let us evaluate $W'(s_0)$. As we know from \eqref{W} and \eqref{nu},
\begin{eqnarray}
    W(s) = 2i\tilde{B} \implies \frac{d W}{d s} = 2i \frac{d \tilde{B}}{d\nu}\frac{d\nu}{d s},
\end{eqnarray}
where,
\begin{eqnarray}
    \frac{d\nu}{d s} = \frac{2 i s_n}{\sqrt{-2V''(x_0)}},
\end{eqnarray}
and we have from \eqref{BT} and \eqref{AP},
\begin{eqnarray}
    \tilde{B} &=& Q^{1/4}(x_1,s)\left[ -i A_+ \frac{\sqrt{2\pi} e^{-i \nu \pi /2}}{\Gamma(-\nu )}\right]e^{i \pi (\nu+1) /4} (4k)^{-(\nu+1)/4} (x_0-x_1)^{-(\nu+1)}e^{i \frac{\sqrt{k}}{2}(x_1-x_0)^2}, \nonumber \\
    &=&Q^{1/4}(x_1,s)\left[ -i \frac{\sqrt{2\pi} e^{-i \nu \pi /2}}{\Gamma(-\nu )}\right]e^{i \pi (\nu+1) /4} (4k)^{-(\nu+1)/4} (x_0-x_1)^{-(\nu+1)}e^{i \frac{\sqrt{k}}{2}(x_1-x_0)^2} \nonumber \\
    &&\times\frac{e^{-i \pi \nu /4} (4k)^{-\nu/4} (x_2-x_0)^{-\nu}e^{i \frac{\sqrt{k}}{2}(x_2-x_0)^2}}{Q^{1/4}(x_2,s)}.
\end{eqnarray}
This gives us
\begin{eqnarray}
    W'(s_n) = 2 i \frac{2 i s_n}{\sqrt{-2V''(x_0)}} i \sqrt{2\pi} e^{\frac{i\pi}{4}}e^{i \frac{\sqrt{k}}{2}(x_2-x_0)^2}e^{i \frac{\sqrt{k}}{2}(x_1-x_0)^2}e^{\frac{-i\pi n}{2}}(4k)^{\frac{-n}{2}-\frac{1}{4}}(x_0-x_1)^{-(n+1)}(x_2-x_0)^{-n}.
\end{eqnarray}
In principle, we can also discuss nonlinearity at the horizon. Nonlinear effects at the horizon during ringdown have been discussed in \cite{Khera_2023}.

\begin{figure}
    \centering
    \includegraphics[width=0.6\linewidth]{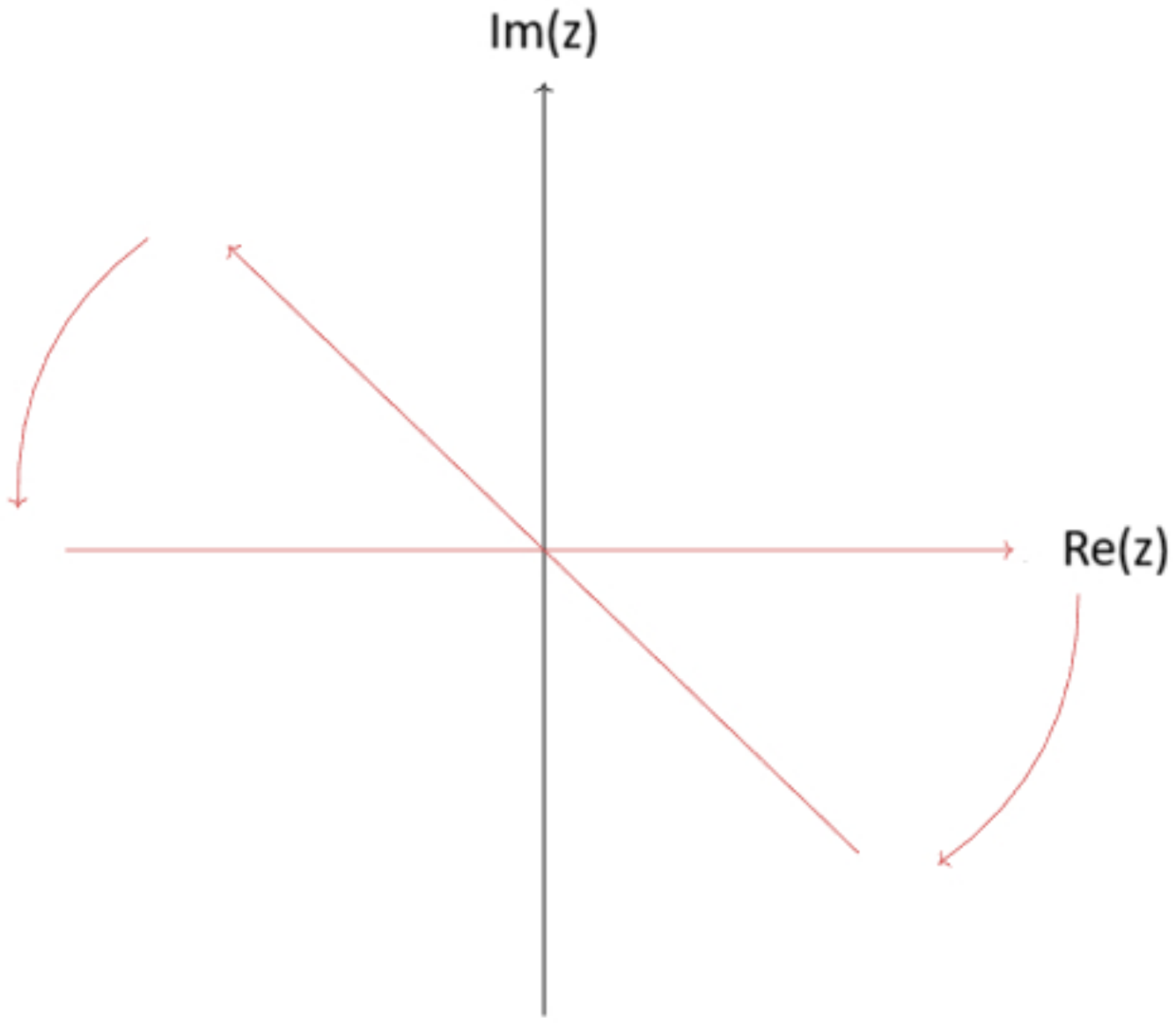}
    \caption{Contour of choice with $z$ being the complexified $(x'-x_0)$}
    \label{contour_of_choice}
\end{figure}

We define, for perturbations computed near the horizon,
\begin{eqnarray}
    NL(x \rightarrow -\infty) &=& \left|\frac{\psi_2}{\psi_1^2}\right| = \left|\frac{1}{2s_0}\frac{\chi}{\psi_1^2}\right|\;\frac{1}{\left|1+\frac{c_{0NL}}{c_{0L}}+\frac{c_w}{c_{0L}}\right|^2}, \nonumber \\
    &\rightarrow& \left|\frac{1}{2s_0}\frac{c_{0{\rm L}}^2 e^{2s_0 T}  \phi_+(x\rightarrow-\infty,2s_0)}{W(2s_0)c_{0L}^2 e^{2s_0 T}\phi_-^2(x\rightarrow -\infty,s_0)}\int_{-\infty}^{+\infty} \; {\rm d}x'\;  \phi_+(x', 2s_0)\; h(x')\right|\;\frac{1}{\left|1+\frac{c_{0NL}}{c_{0L}}+\frac{c_w}{c_{0L}}\right|^2}, \nonumber\\
    &=&\left|\frac{1}{2s_0}\sqrt{\frac{s_0}{2}}\frac{1}{W(2s_0)}\frac{1}{\left|1+\frac{c_{0NL}}{c_{0L}}+\frac{c_w}{c_{0L}}\right|^2}\int_{-\infty}^{+\infty} \; {\rm d}x'\;  \phi_+(x', 2s_0)\; h(x')\right|.
\end{eqnarray}
Here, we use the near-horizon behaviour of the solutions as derived in \eqref{phi+}, \eqref{I1 horizon}, to simplify the expression.
Let us now look at the integral, using the solution we have for $\phi_+(x,s)$ from \eqref{phi+}, and applying the method of steepest descent about the maximum $x_0$, to bring $H(x_0)$ outside the integral.
\begin{eqnarray}
    &&\int_{-\infty}^{+\infty} \; {\rm d}x'\;  \phi_+(x', 2s_0)\; h(x'), \nonumber \\
    &=&\int_{-\infty}^{+\infty} \; {\rm d}x'\;  A_+(2s_0)D_{\nu}(t')A^2_-(s_0)D^2_0(t')\; H(x'), \nonumber \\
    &=&H(x_0)A_+(2s_0)A^2_-(s_0)\frac{2^{\frac{\nu}{2}}\sqrt{\pi}}{\Gamma(\frac{1-\nu}{2})}\int_{-\infty}^{+\infty} \; {\rm d}x'e^{\frac{-3i\sqrt{k}}{2}(x'-x_0)^2}, \nonumber \\
    &=&H(x_0)A_+(2s_0)A^2_-(s_0)\frac{2^{\frac{\nu}{2}}\sqrt{\pi}}{\Gamma(\frac{1-\nu}{2})}\sqrt{\frac{2\pi}{3i\sqrt{k}}}.
\end{eqnarray}
The Gaussian integral has been computed by rotating the contour, as shown in \ref{contour_of_choice}.
Now, for our calculations, we shall use the source term for the $l=2$, $m=0$ mode for the channel $(2,0) \times (2,0) \to (2,0)$. This source term is taken from \cite{Gleiser_1996}, and we have corrected one term in that source following footnote 1 in \cite{Gleiser_2000}. We use the corrected Laplace transformed source term in our calculations. Denoting $\psi_r = \frac{d\psi}{dr}$ and $\psi_{rr} = \frac{d^2 \psi}{dr^2}$, the source is
\begin{eqnarray}
\label{Sriotto}
S_{l=2,\,m=0}(r,T)=e^{2s_0 T} \;\frac{12 (r-2M)^3}{7 (2 r+3M)}\nonumber
\left[\begin{split}
&\frac{(r-2M) (2 r+3M) \psi_{rr}^2}{3 r^4}-\frac{s_0 M \psi  \psi_{rr}}{r^3 (2 r+3M)}+\frac{(3 r-7M) s_0 \psi _r^2}{3 (r-2M) r^3}+\frac{s_0 \psi _r \psi_{rr}}{3 r^2} \nonumber \\
&-\frac{s_0 \psi  \psi_{rrr}}{3 r^2}
-\frac{(2 r+3M) s_0^2 \psi _r^2}{3 (r-2M) \
r^2}+\frac{4 \left(3 r^2+5 r M+6M^2\right) \psi _r \psi_{rr}}{3 r^5}\nonumber \\
&-\frac{12 \left(r^2+M r+M^2\right)^2 \
s_0^2 \psi ^2}{(r-2M)^3 r^4 (2 r+3M)}+\frac{\left(8 r^2+12 r M+7M^2\right) s_0 \psi  \psi _r}{(r-2M) r^4 (2 \
r+3M)}\nonumber \\
&-\frac{\left(2 r^2-M^2\right) s_0 \psi  \psi _{rr}}{(r-2M) r^3 (2 r+3M)}-\frac{4 \
\left(r^2+r M+M^2\right) s_0^2 \psi  \psi _r}{(r-2M)^2 r^3}\nonumber \\
&-\frac{4 \left(2 r^3+4 r^2 M+9 r M^2+6M^3\right) \psi  \psi_{rr}}{r^6 (2 r+3M)}+
\frac{\left(12 r^3+36 r^2 M+59 r M^2+90 M^3\right) \psi _r^2}{3 (r-2M) \
r^6}\nonumber \\
&+\frac{\left(18 r^3-4 r^2 M-33 r M^2-48 M^3\right) s_0 \psi  \psi _r}{3 (r-2M)^2 r^4 (2 r+3M)}+\\
& \frac{\left(112 \
r^5+480 r^4 M+692 r^3 M^2+762 r^2 M^3+441 r M^4+144 M^5\right) s_0 \psi ^2}{(r-2M)^2 r^5 (2 r+3M)^3}+ \nonumber \\
& \frac{12 \left(2 r^5+9 \
r^4 M+6 r^3 M^2-2 r^2 M^3-15 r M^4-15 M^5\right) \psi ^2}{(r-2M)^2 r^8 (2 r+3M)}\nonumber \\
&-\frac{2 \left(32 r^5+88 r^4 M+296 r^3 M^2+510 \
r^2 M^3+561 r M^4+270 M^5\right) \psi  \psi _r}{(r-2M) r^7 (2 r+3M)^2}\end{split}\right],
\end{eqnarray}
where $\psi$ is the linear $l=2$, $m=0$ mode.\vspace{2mm}\\
Now all that is left is to calculate the coefficients $c_{nL}$, $c_{nNL}$, $s_0$ and $\nu$, for which we need to choose matching points $x_1$ and $x_2$ in the allowed range.

Now let us find $s_0$, the most dominant QNM mode. For the WKB approximation, we get this by taking $n=0=\nu$.
We know from \eqref{nu},
\begin{eqnarray}
    \nu+\frac{1}{2} = i \frac{[s^2+V(x_0)]}{\sqrt{-2 V''(x_0)}}.
\end{eqnarray}
Putting $\nu=0$
\begin{eqnarray}
    \frac{1}{2} = i \frac{[s_0^2+V(x_0)]}{\sqrt{-2 V''(x_0)}}
\end{eqnarray}
\begin{eqnarray}
    &\implies s_0^2 + V(x_0) = -\frac{i\sqrt{-2V''(x_0)}}{2},\nonumber\\
    &\implies s_0^2 = -V(x_0)-\frac{i\sqrt{-2V''(x_0)}}{2}
    &\implies s_0 = -( 0.088-0.398i).
\end{eqnarray}
Here we choose the sign of $s_0$ that has a negative real part, as the oscillations are decaying with time. Next, we move on to calculating the coefficients $c_{nL}$ and $c_{nNL}$. As observed in \cite{Perrone_2024}, the nonlinearity of the channel we are considering in this section is sensitive to initial conditions. So we have to choose initial conditions corresponding to a head-on collision of black holes which will excite the $(2,0)$ mode.

Following \cite{Perrone_2024}, for this we refer to the Misner wormhole solutions as the initial conditions \cite{Misner}. This corresponds to two black holes very close to each other at a distance $L$ and with initial momentum $P$.
\begin{eqnarray}
\psi_{T=0} &=& f_1(x) =
{\displaystyle \frac {8}{3}}{\displaystyle
\frac {{M}{L}^{2} \left(  5\sqrt {{r} - 2{M}} + 7
\sqrt {{r}}   \right) {r}}{ \left(  \sqrt {{r}} +
\sqrt {{r} - 2{M}}   \right) ^{5}(2{r} + 3{M})}}, \label{psit0}\\
\dot{\psi}_{T=0} &=& g_1(x) =
{\displaystyle \frac {\sqrt{{r} - 2{M}}
{P}{L}(8
{r} + 6{M})}{{r}^{5/2}(2{r} + 3{M})}}, \\
\chi_{T=0} &=& f_2(x) =- {\displaystyle \frac {512}{7}}
{\displaystyle \frac {{M}^{2}{L}^{4}}{ \left(  \sqrt {{r}}
 + \sqrt {{r} - 2{M}}   \right) ^{10}{r}}} +
{\displaystyle \frac {16}{7}}{\displaystyle
\frac { \left(  9\sqrt {{r}} +
17\sqrt {{r} - 2{M}}   \right) \sqrt {r - 2{M}}{M}{P}{L}^{3}}{
 \left(  \sqrt {{r}} + \sqrt {{r} - 2{M}}   \right) ^{5
}
(2{r}
 + 3{M}){r}^{5/2}}}, \\
\dot{\chi}_{T=0} &=& g_2(x) =
{\frac {64\,{M}^{2}\left (10\,r-10\,M+38\,\sqrt {r}\sqrt {r-2\,M}
\right )\sqrt {r-2\,M}{L}^{4}}{7\,\left (\sqrt {r}+\sqrt {r-2\,M}
\right )^{10}\left (4\,r+6\,M\right ){r}^{5/2}}} - \nonumber \\
&-& \,{\displaystyle \frac {64}{7}}\,
{\displaystyle \frac {(\,4\,{r} + 14\,{M}\,)\sqrt{r-2M}\,{M}\,{P}\,{L}^{3}}{
 \left( \! \,\sqrt {{r}} + \sqrt {{r} - 2\,{M}}\, \!  \right) ^{5
}\,{r}^{5}}}
- 64\,{\displaystyle \frac {{M}\,
 \left( \! \,5\,\sqrt {{r}} - 3\,\sqrt {{r} - 2\,{M}}\, \!
 \right) \,\sqrt {{r} - 2\,{M}}\,{L}^{2}\,{P}^{2}\,{\it q2}}{
 \left( \! \,\sqrt {{r}} + \sqrt {{r} - 2\,{M}}\, \!  \right) ^{5
}\,(\,2\,{r} + 3\,{M}\,)\,{r}^{5/2}}} \nonumber\\
 &&+
{\displaystyle \frac {16}{35}} \sqrt {{r} - 2\,{M}}\,
{L}^{2}\,{P}^{2} \left( 1750 M^4-9849rM^3+2331r^2M^2+7182r^3M-2892 r^4 \right. \nonumber \\
&&
\left.
-\sqrt{r} \sqrt{r-2M} (3148 r^3-4130r^2M-4935rM^2+4375M^3)
 \right) \left/ \right. \! \!  \left( \! \,(\,2\,{r} + 3\,{M}\,)\, \left( \! \,
\sqrt {{r}} + \sqrt {{r} - 2\,{M}}\, \!  \right) ^{5}\,{r}^{6}\,
 \!  \right),  \nonumber\label{chidott0}\\
\end{eqnarray}
where we shall use the values $P=0$ and $L=2.1$\\ \vspace{2mm}
Using these initial conditions, we can evaluate $c_{nL}$ and $c_{nNL}$ using \eqref{cnL} and \eqref{cnNL}.
\begin{eqnarray}
    c_{0L} &=& \frac{1}{W'(s_0)}\int_{-\infty}^{+\infty} \; {\rm d}x'\;  \, \phi_-(x', s_0)\; \left( -s_0\,f_1(x') - g_1(x') \right), \nonumber \\
    &=& \frac{A_-(s_0)}{W'(s_0)}\int_{-\infty}^{+\infty} \; {\rm d}x'\;  \, D_0(t')\; \left( -s_0\,f_1(x') - g_1(x') \right), \nonumber \\
    &=& \frac{A_-(s_0)}{W'(s_0)}\int_{-\infty}^{+\infty} \; {\rm d}x'\;  \, e^{\frac{-i\sqrt{k}}{2}(x-x_0)^2}\; \left( -s_0\,f_1(x') - g_1(x') \right), \nonumber \\
    &=& \frac{A_-(s_0)}{W'(s_0)}\left( -s_0\,f_1(x_0) - g_1(x_0) \right)\sqrt{\frac{2\pi}{i\sqrt{k}}}.
\label{c0L}
\end{eqnarray}
Similarly replacing $f_1(x)$ and $g_1(x)$ with $f_2(x)$ and $g_2(x)$ gives us $c_{0NL}$.\\
Now, we want the nonlinearities of the gravitational strain $h$
at infinity. First, we note that the WKB functions we used have the conventional WKB normalization with a factor of $\sqrt{\omega}$ in the denominator as in, for example, (\ref{I3}).
So, to convert to the standard normalization of QNMs, we
multiply, for example,  $\phi_{\pm} (x, 2s_0)$ by a factor $\sqrt{2\omega}$ and $\phi_{-}(x, s_0)$ by a factor $\sqrt{\omega}$ in (\ref{NLI}). This gives an overall factor $(2\omega)$ multiplying our original nonlinearity ratio.
Now, the strain is directly related to the angular metric perturbation in TT gauge. However, our perturbations are in Regge-Wheeler gauge at first order. We have used the same gauge at second order, and regularized the source. The Zerilli function is directly related to the angular metric perturbation as can be seen, for example in (\ref{so10}).
In order to compute the strain nonlinearity ratio, we have to make a gauge transformation to TT gauge. Fortunately, this has already been done in section 4.2 of \cite{bucciottiamplitudes}.
The angular metric perturbation in our gauge is $(i\omega)$ $\times$ (strain amplitude). Thus, to reconstruct the strain nonlinearity ratio $NL_h$ from our nonlinearity ratio $NL_{\psi}$, including the conversion from WKB normalization, we have the following conversion factor:
\begin{eqnarray}
NL_{h} &=& NL_{\psi} \times \left| (2 \omega) (\frac{1}{2i\omega}
) (i\omega)^2 \right|, \nonumber \\
&=& NL_{\psi} |\omega^2 |.
\label{conversion}
\end{eqnarray}
Putting all this in the formulae for nonlinearity gives us
\begin{eqnarray}
    NL_{\psi, l=2}(\infty)= 98362.5,
\end{eqnarray}
and
\begin{eqnarray}
    NL_{h, l=2}(\infty)= 16418.9.
\end{eqnarray}

\begin{figure}
    \centering
    \includegraphics[width=0.60\linewidth]{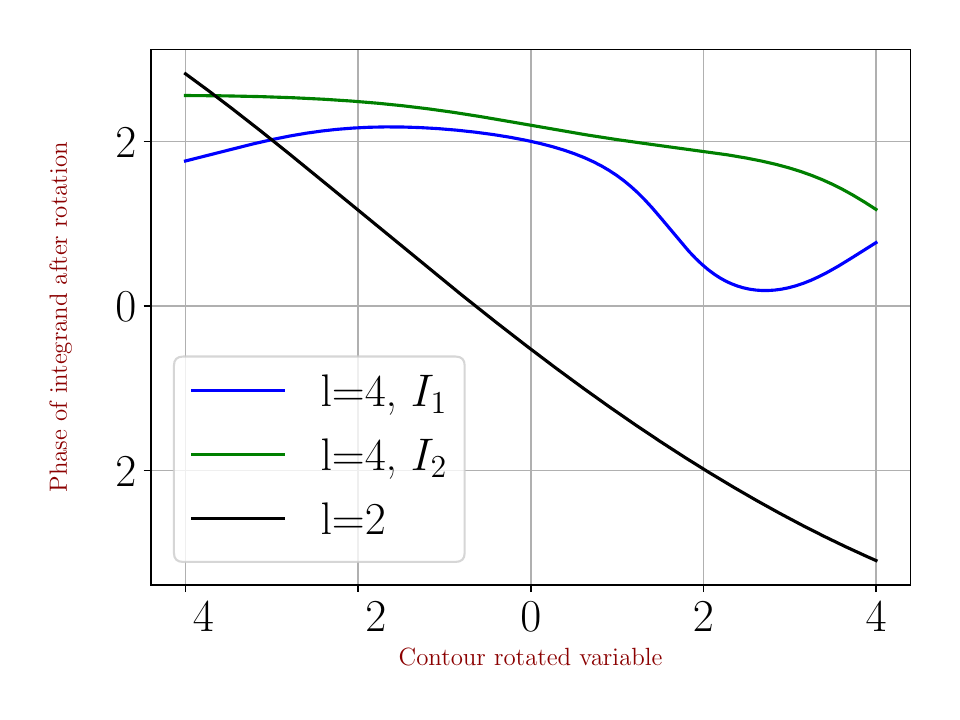}
    \caption{x-axis: contour rotated variable, with $x=0$ being the point about which we apply the saddle-point approximation\\
    y-axis: Black line shows the phase of the integrand for $l=2$, blue and green lines for $l=4$}
    \label{phaseplot}
\end{figure}
We see that the nonlinearity ratio at infinity for $l=2$ comes out through this method to be a very large value. This is certainly not physical. While the nonlinearity ratio for this channel has not been discussed in the available numerical simulations, the nonlinearity ratio is not this high for any of the channels studied. Why is the number obtained so large? This is because the method of steepest descent is not applicable for the naive choice of contour.  For the contour we use to integrate the imaginary Gaussian in \ref{imgaussian}, the full integrand does not have stationary phase along the contour. The contour that must be chosen for integrating the imaginary Gaussian in \ref{imgaussian} is Figure \ref{contour_of_choice}.

We see that the contribution of the integrand along the arcs at positive and negative infinity goes to zero.
The integral along the real line is equal modulo a constant to the integral along the diagonal line. Along the diagonal line, the coordinate $x$ is such that
\begin{eqnarray}
    x = (x'-x_0)e^{\frac{3i\pi}{4}},
\end{eqnarray}
and thus we just have to evaluate the integral for $(x' - x_0)$ running from minus infinity to plus infinity. While the imaginary Gaussian becomes a real Gaussian along this contour and thus has constant phase, the source contributions coming from $H(x)$ do not have a stationary phase along this contour, thus invalidating the use of method of steepest descent with this choice of contour in the $l=2$\ case.
This is seen in Figure \ref{phaseplot}, where the phase of the entire integrand is plotted in the black curve. We are interested in a small region about $x=0$ in this plot, where the spatially truncated QNMs are expected to be non-zero. The phase of the integrand is certainly not stationary in a neighborhood of $x=0$.
A similar problem arises when evaluating $c_W$ through the saddle point approximation in this case. We can also use this method in principle, to calculate the nonlinearity ratio near the horizon, as done by \cite{Khera_2023}. However, the method of steepest descent is not valid in this case as well for our contour. To find the exact contour of stationary phase is not feasible as our integrand is a complicated expression with parabolic cylinder functions.\\

This explains the large value for nonlinearity obtained in the steepest descent approximation. This, however is not the case for other values of $l>2$ where the integrand after rotation has stationary phase along the contour of choice for the region which contributes to the integral. This will be discussed in the next section. \\

When the steepest descent method fails for this contour, we don't know the exact values of $x$ for which we should consider the integrand non-zero and perform the integral. This is because, as discussed before, causality leads to spatially truncated QNMs in the integral. However, we do not have a precise range over which they are non-zero. That range will in general, depend on the potential and also the support of the initial data (it roughly lies inside the common future light cone of the initial data \cite{szpak2004quasinormalmodeexpansionexact}). Let us perform numerical integration in Mathematica (rather than steepest descent) to evaluate the value of the relevant integral between the matching points, taking them to be the turning points. We then indeed get a \textit{much} smaller nonlinearity ratio (than the steepest descent approximation) for the gravitational strain
\begin{eqnarray}
    NL_{l=2, N-integrate}(\infty)= 0.313,
\end{eqnarray}
indicating further evidence for our explanation. In this computation, we have numerically integrated all the integrals needed in the computation of the nonlinearity ratio between the approximate turning points.
We do not know the range of $x$ over which the spatially truncated QNMs are non-zero, and integrals over different ranges can in principle,  lead to different nonlinearity ratios. Thus, for this channel, we cannot reliably calculate the nonlinearity ratios in an analytical approximation --- we can just get a rough estimate by integrating between the matching points.
\section{Nonlinearity ratio for $l=4$}\label{sec10}
In this section we shall proceed to evaluate the non-linearities for the $(2,\pm2)\times(2,\pm2)\rightarrow(4,4)$ channel. Here, we can successfully reproduce nonlinearity ratios seen in simulations, as the method of steepest descent is valid in this case.

Until now, we have calculated the nonlinearity for $l=2$ mode made by $(2,0)$ and $(2,0)$ modes and thus the Gaussian factor in $\phi_+(x,2s_0)$ and the Gaussian factor in $\phi_-(x,s_0)$ are both peaked at $x_0$, the peak of the $l=2$ potential. However, this will not be the case for  $l>2$ modes made of $l=2$ and $l=2$ modes, as the maximum of the $l>2$ Zerilli potential will not be at the same point as that for the $l=2$ case. Let us thus perform this integral for a general $l_2\neq l_1$ mode made out of $l=l_1$ and $l=l_1$ modes. We have from \eqref{NLI},
\begin{eqnarray}
    &&NL(x \rightarrow \infty) = \left|\frac{\psi_2}{\psi_1^2}\right| = \left|\frac{1}{2s_0}\frac{\chi^S}{\psi_1^2}\right|, \nonumber \\
    &&=\left|\frac{1}{2s_0}\frac{\phi_+(x\rightarrow\infty,2s_0)}{\phi_-^2(x\rightarrow \infty,s_0)}\frac{1}{W(2s_0)}\int_{-\infty}^{\infty}\phi_-(x,2s_0)\phi_-^2(x,s_0)H(x){\rm d}x\right|, \nonumber \\
    &&=\left|\frac{1}{2 s_0}\sqrt{\frac{s_0}{2}}\frac{1}{W(2s_0)}\frac{1}{C^2}\int_{-\infty}^{\infty}\phi_-(x,2s_0)\phi_-^2(x,s_0)H(x){\rm d}x\right|.
\end{eqnarray}
We have the integral for nonlinearity at infinity, along with the solution for $\phi_-(x,s)$ from \eqref{phi-}
\begin{eqnarray}
    &&\int_{-\infty}^{\infty}\phi_-(x,2s_0)\phi_-^2(x,s_0)H(x){\rm d}x \nonumber \\
    &=&\int_{-\infty}^{\infty}[aD_{\nu}(t)+bD_{-1-\nu}(it)]\phi_-^2(x,s_0)H(x){\rm d}x   .
\end{eqnarray}
Let us name
\begin{eqnarray}
    I_1 = \int_{-\infty}^{\infty}[aD_{\nu}(t)]\phi_-^2(x,s_0)H(x){\rm d}x  ,
\end{eqnarray}
and
\begin{eqnarray}
    I_2 = \int_{-\infty}^{\infty}[bD_{-1-\nu}(it)]\phi_-^2(x,s_0)H(x){\rm d}x .
\end{eqnarray}
Now,
\begin{eqnarray}
    &&I_1= \int_{-\infty}^{\infty}{\rm d}x\, a\; A_-^2 H(x) e^{-i\sqrt{k_{l_1}}(x-z_1)^2}e^{-i\frac{\sqrt{k_l}}{2}(x-z_2)^2}\left[\frac{2^{\frac{\nu}{2}}\sqrt{\pi}}{\Gamma(\frac{1-\nu}{2})}\;{}_1F_1(\frac{-\nu}{2},\frac{1}{2},\frac{{t(x)}^2}{2})-\frac{\sqrt{2 \pi}2^{\frac{\nu}{2}}}{\Gamma(-\frac{\nu}{2})}{t(x)}\;{}_1F_1(\frac{1-\nu}{2},\frac{3}{2},\frac{{t(x)}^2}{2})\right], \nonumber\\
    &&=\int_{-\infty}^{\infty}{\rm d}x\, a\; A_-^2 H(x) \left[\frac{2^{\frac{\nu}{2}}\sqrt{\pi}}{\Gamma(\frac{1-\nu}{2})}\;{}_1F_1(\frac{-\nu}{2},\frac{1}{2},\frac{{t(x)}^2}{2})-\frac{\sqrt{2 \pi}2^{\frac{\nu}{2}}}{\Gamma(-\frac{\nu}{2})}{t(x)}\;{}_1F_1(\frac{1-\nu}{2},\frac{3}{2},\frac{{t(x)}^2}{2})\right], \nonumber \\
    &\times& e^{-i\frac{\alpha \beta}{\alpha+\beta}(z_1-z_2)^2} e^{-i(\alpha+\beta)(x-x_{m1})^2}, \nonumber
\end{eqnarray}
where $\alpha=\frac{\sqrt{k_{l_2}}}{2}$, $\beta = \sqrt{k_{l_1}}$ and $z_2 = x_{l_2}$ and $z_1 = x_{l_1}$ represent the maxima of the potentials corresponding to $l_2$ and $l_1$ respectively. We thus see that the exponential term after contour rotation has its maximum at
\begin{eqnarray}
    x_{m1} = \frac{\alpha z_2+\beta z_1}{\alpha+\beta}.
\end{eqnarray}
We have just used the fact that a product of Gaussians peaked about two different points is a Gaussian peaked about the point $x_{m1}$.
We shall apply the method of steepest descent about this point. Performing the contour integral, with $\alpha+\beta>0$ gives us
\begin{eqnarray}
    &&I_1 = a A_-^2 e^{-i\frac{\alpha \beta}{\alpha+\beta}(z_1-z_2)^2}  H(r_{m1}), \nonumber\\
    &&\times \left[\frac{2^{\frac{\nu}{2}}\sqrt{\pi}}{\Gamma(\frac{1-\nu}{2})}\;{}_1F_1(\frac{-\nu}{2},\frac{1}{2},\frac{{t(x_{m1})}^2}{2})-\frac{\sqrt{2 \pi}2^{\frac{\nu}{2}}}{\Gamma(-\frac{\nu}{2})}{t(x_{m1})}\;{}_1F_1(\frac{1-\nu}{2},\frac{3}{2},\frac{{t(x_{m1})}^2}{2})\right]\sqrt{\frac{\pi}{i(a_1+b_1)}}.
\end{eqnarray}
Here, $r_{m1}$ is the value of $r$ corresponding to the tortoise coordinate $x_{m1}$.
Next,
\begin{eqnarray}
    &&I_2 = b A_-^2 e^{i\frac{\alpha \beta}{\beta-\alpha}(z_1-z_2)^2}e^{-i(-\alpha+\beta)(x-\frac{-\alpha z_2+\beta z_1}{-\alpha+\beta})^2}H(x), \nonumber\\
    &&\times \left[\frac{2^{\frac{-1-\nu}{2}}\sqrt{\pi}}{\Gamma(1+\frac{\nu}{2})}\;{}_1F_1(\frac{1+\nu}{2},\frac{1}{2},\frac{-{t(x)}^2}{2})-\frac{\sqrt{2 \pi}2^{\frac{-1-\nu}{2}}}{\Gamma(\frac{1+\nu}{2})}{it(x)}\;{}_1F_1(1+\frac{\nu}{2},\frac{3}{2},\frac{-{t(x)}^2}{2})\right].
\end{eqnarray}
Here, we have $\beta-\alpha<0$. Let us now apply the method of steepest descent about the maximum of the resultant Gaussian, which is at
\begin{eqnarray}
    x_{m2} = \frac{-\alpha z_2+\beta z_1}{-\alpha+\beta}.
\end{eqnarray}
This gives us
\begin{eqnarray}
    &&I_2 = b A_-^2 e^{i\frac{\alpha \beta}{\beta-\alpha}(z_1-z_2)^2} \nonumber H(r_{m2}) \nonumber \\ &&\left[\frac{2^{\frac{-1-\nu}{2}}\sqrt{\pi}}{\Gamma(1+\frac{\nu}{2})}\;{}_1F_1(\frac{1+\nu}{2},\frac{1}{2},\frac{-{t(x_{m2})}^2}{2})-\frac{\sqrt{2 \pi}2^{\frac{-1-\nu}{2}}}{\Gamma(\frac{1+\nu}{2})}{it(x_{m2})}\;{}_1F_1(1+\frac{\nu}{2},\frac{3}{2},\frac{-{t(x_{m2})}^2}{2})\right]\sqrt{\frac{\pi}{i(b_1-a_1)}}.
\end{eqnarray}
We have the Laplace transform of the source term for $l=4$ mode, computed in \cite{Nakano_2007}:

\begin{align*}
\label{Snakano}
S_{l=4,\,m=4}(r,T)=-e^{2s_0 T}\;\frac{1}{9} \sqrt{\frac{5}{14 \pi }} s_0 \Bigg(\Bigg(\frac{(r-2M) (7 r+4M) \
s_0^2}{r}-\frac{3 (r-2M)}{r^5 (2 r+3M)^2 (3 r+M)^2} (228 r^7+ \nonumber \\
+8 r^6M-370 r^5M^2+142 r^4M^3-384 r^3M^4-514 r^2M^5-273 rM^6-48M^7)
\Bigg) \psi_r^2 \nonumber \\
+\psi \psi_r  \Bigg(\frac{4 M^2(r-2M)^2 s_0^2}{r^2 (3 r+M)^2}+\frac{6 (r-2M)}{r^6 (2 r+3M)^3 (3 r+M)^2} (144 r^8+4116 r^7M+2154 r^6M^2-2759 r^5M^3-8230 r^4M^4- \nonumber \\
9512 r^3M^5-3540 r^2M^6-1119 rM^7-144M^8)\Bigg) \nonumber \\
+\psi^2 \Bigg(-\frac{3 (276 r^7+476 r^6M-1470 r^5M^2-1389 r^4M^3-816 r^3M^4-800 r^2M^5-555 rM^6-96M^7) s_0^2}{(r-2M) r^3 (2 r+3M)^2 (3 r-M)^2}+
\nonumber \\
\frac{9 (r-2M)}{r^7 (2 r+3M)^4 (3 r+M)^2} (2160 r^9+11760 r^8M+30560 r^7M^2+ \nonumber \\
41124 r^6M^3+31596 r^5M^4+11630 r^4M^5-1296 r^3M^6-4182 r^2M^7-1341 rM^8-144M^9)-\frac{r (7 r+4M) s_0^4}{r-2M}\Bigg)
\Bigg).
\end{align*}
Using this source term in the expression for nonlinearity gives us the nonlinearity for the perturbations,
\begin{eqnarray}
    NL_{l=4,\psi}(x \rightarrow \infty) = 0.981,
\end{eqnarray}
and including the extra factor $s^2_0$ in the numerator following from \eqref{conversion} gives us the value of nonlinearity for gravitational strain as
\begin{eqnarray}
    NL_{l=4, h}(x \rightarrow \infty) = 0.1638.
\end{eqnarray}
This matches perfectly, numerical values obtained by \cite{Cheung_2023},\cite{Bucciotti_2024},\cite{Redondo_Yuste_2024}. We see that the linear $(2, \pm2)$ mode amplitude does not get renormalized, since the nonlinear contributions from $c_W$ and $c_{0NL}$ are at the lowest QNM frequency of the $l=4$ potential, whereas the parent linear mode has QNM frequency corresponding to the $l=2$ potential. However, the amplitude of the $l=2$ mode could get renormalized from other channels that are also excited by the merger. This is not expected to be very significant, hence we neglect it. As we see, there is still a good match to numerical values of the nonlinearity ratio. Finally, we also plot the phase of the integrands in the integrals $I_1$ and $I_2$, which are the blue and green curves in Figure \ref{phaseplot}. We see that in a neighborhood of $x=0$, the phase does not vary much. Thus, the method of steepest descent is a reliable approximation.

Recently, the authors of \cite{Khera_2023} have found evidence of a QQNM at the horizon by studying simulations of head-on collisions of non-spinning black holes. They have also computed nonlinearity ratios. Their nonlinearity ratios are for $m=0$, whereas we have $m \neq 0$.
Motivated by their result, we can calculate the nonlinearity ratio near the horizon. We have checked that the method of steepest descent is applicable.

This gives the nonlinearity ratio for the perturbations as
\begin{eqnarray}
    NL_{l=4, \psi}(x \rightarrow - \infty) = 0.330.
\end{eqnarray}

The results for \cite{Khera_2023} are for the ratios of shear amplitudes, which are the natural quantities to study at the horizon. On the horizon, the shear is
the tracefree part of $\frac{1}{2} \nabla_v h_{AB}$, where $A,B$ are coordinates on the transverse cross-section of the horizon and $v$ is the Eddington-Finkelstein ingoing coordinate. The ratio of shear amplitudes gives a factor $4/|\omega|$ multiplying the above nonlinearity ($\omega$ is the $l=2$ lowest QNM). The WKB normalization means that we have to multiply by an overall factor of $2\omega$. We therefore get the nonlinearity amplitude of shear ratios to be
\begin{equation}
NL_{l=4, ~shear}(x \rightarrow - \infty) = 2.64.
\end{equation}
While \cite{Khera_2023} study shear nonlinearity ratios for channels different from us, their results for the channels
$(2,0) \times (2,0) \to (2,0)$ and $(2,0) \times (2,0) \to (4,0)$ are $1.51$ and $0.73$ respectively, which are of the same order. Our results also give clear evidence of nonlinear effects at the horizon.
    \subsection{Gauge dependence in source}
It is well-known that if we use Regge-Wheeler gauge, which is not asymptotically flat, the source term at second-order blows up at infinity and needs to be regularized \cite{Nakano_2007}. We have used the regularized source in \cite{Nakano_2007} where the following term is subtracted from $\chi$ in the second-order Zerilli equation to regularize the raw source term:
\begin{eqnarray}
     \chi_0 = \frac{1}{126}\sqrt{\frac{70}{\pi}}\frac{(r-2)^2}{r}\psi_{,r}\psi_{,rT}.
\end{eqnarray}
However, as noted in \cite{Nakano_2007}, this choice is not unique. We could replace $\frac{\partial}{\partial r}$ by $-\frac{\partial}{\partial T}$, by subtracting from $\chi$ instead, the term
\begin{eqnarray}
    \chi'_0 =  \frac{1}{126}\sqrt{\frac{70}{\pi}}\left[\frac{(r-2)^2}{r}+\frac{16}{r^3}\right]\psi_{,T}\psi_{,TT}.
\label{regterm}
\end{eqnarray}
A priori, it is not clear that it will lead to the same nonlinearity ratio the way we have calculated it. We have used the method of steepest descent at a point far away from infinity, the nonlinearity explicitly depending on the source term in the region near the maximum. Using \ref{regterm} to regularize the source instead, we get
\begin{eqnarray}
NL_{l=4}(x \rightarrow \infty) = 0.1668,
\end{eqnarray}
which is close to the value we calculated using the original regularized source choice. We list a few more choices of regularization in Table \ref{nl table}, and see that they do not change the nonlinearity ratio appreciably. Some other choices lead to divergence of the source term at the horizon, and therefore we have not considered them. For the same choices of regularization, we find the nonlinearity ratio at the horizon also does not change appreciably as seen in Table \ref{nl table horizon}. The source terms for the various choices in the table are given in the appendix.
\begin{table}[h!]
\centering
\begin{tabular}{l c c}
\hline
\textbf{$\chi'_0$} & \textbf{$NL_{l=4}(x \rightarrow \infty)$} \\
\hline
$\frac{1}{126}\sqrt{\frac{70}{\pi}}\left[\frac{(r-2)^2}{r}+\frac{16}{r^3}\right]\psi_{,T}\psi_{,TT}$ & 0.1668 \\
$\frac{1}{126}\sqrt{\frac{70}{\pi}}\left[\frac{(r-2)^2}{r}+\frac{16}{r^3}\right]\psi_{,x}\psi_{,xT}$ & 0.1645 \\
$\frac{1}{126}\sqrt{\frac{70}{\pi}}\left[-\frac{(r-2)^2}{r}+\frac{16}{r^3}\right]\psi_{,T}\psi_{,xT}$ & 0.1641 \\
\hline
\end{tabular}
\caption{Nonlinearity at infinity --- dependence on gauge}
\label{nl table}
\end{table}
\begin{table}[h!]
\centering
\begin{tabular}{l c c}
\hline
\textbf{$\chi'_0$} & \textbf{$NL_{l=4}(x \rightarrow -\infty)$} \\
\hline
$\frac{1}{126}\sqrt{\frac{70}{\pi}}\left[\frac{(r-2)^2}{r}+\frac{16}{r^3}\right]\psi_{,T}\psi_{,TT}$ & 0.056 \\
$\frac{1}{126}\sqrt{\frac{70}{\pi}}\left[\frac{(r-2)^2}{r}+\frac{16}{r^3}\right]\psi_{,x}\psi_{,xT}$ & 0.056 \\
$\frac{1}{126}\sqrt{\frac{70}{\pi}}\left[-\frac{(r-2)^2}{r}+\frac{16}{r^3}\right]\psi_{,T}\psi_{,xT}$ & 0.055 \\
\hline
\end{tabular}
\caption{Nonlinearity at horizon --- dependence on gauge}
\label{nl table horizon}
\end{table}
\subsection{Sensitivity to choice of matching point}
We have already discussed the allowed range of matching points in Section \ref{sec7}.
The nonlinearity has a negligible dependence on the choice of $x_1$, $x_2$, the matching points at the linear order and $y_1,y_2$ the matching points at second order.

Varying the matching points in the allowed range gives us the values of nonlinearity in Table \eqref{nl table x1}. In the first row are the matching points at the extreme left end of the range, the turning points. We see that the nonlinearity ratio does not change much. The only thing we have to be careful about is to keep the width of the matching region the same at first and second order (roughly the same split of three regions), otherwise the ratio changes. This appears reasonable, since the nonlinearity ratio involves an integral over a product of linear and second-order mode solutions in the same middle region.
\begin{table}[h!]
\centering
\begin{tabular}{l c c}
\hline
\textbf{$x_0-x_1$} & \textbf{$x_4-y_1$} & \textbf{$NL_{l=4}(x \rightarrow \infty)$} \\
\hline
3.766 & 3.478 & 0.164 \\
8 & 8 & 0.165 \\
9 & 9 & 0.163 \\
10 & 10 & 0.162 \\
\hline
\end{tabular}
\caption{Nonlinearity at infinity --- dependence on matching points}
\label{nl table x1}
\end{table}\\

\begin{table}[h!]
\centering
\begin{tabular}{l c c}
\hline
\textbf{$x_0-x_1$} & \textbf{$x_4-y_1$} & \textbf{$NL_{l=4}(x \rightarrow \infty)$} \\
\hline
3.766 & 3.478 & 0.055 \\
8 & 8 & 0.055 \\
9 & 9 & 0.055 \\
10 & 10 & 0.054 \\
\hline
\end{tabular}
\caption{Nonlinearity at horizon --- dependence on matching points}
\label{nl horizon table x1}
\end{table}

Here, $x_0,x_4$ are the maxima of the $l=2,4$ potential respectively, while $y_1,y_2$ are the matching points for $2s_0$ frequency. The first row thus corresponds to $x_1,x_2$ and $y_1,y_2$ being the linear and second-order turning points. We choose our matching points to be close by at the linear and nonlinear order so that we have roughly the same split of three regions.

\section{Adding overtones to the analysis}\label{sec11}
All the analysis so far has been by considering only the dominant linear QNM corresponding to some $(l,m)$. Higher overtones decay faster and will not be as significant as the dominant mode. However, for a more precise calculation, we can also compute the overtone contribution using the same techniques. At the end of this section, we will also mention one application of computing overtone nonlinearity ratios.
As we know from \eqref{psi}, the linear perturbation is given as a sum over the various QNMs $s_n$ as,
\begin{align}
\psi(x,T) = \sum_{n}c_{nL}e^{Ts_n}\phi_{-}(x,s_n).
\end{align}
This perturbation sources the second order perturbation.

Solving the second order Zerilli equation gives us three contributions as before, to the second order solutions. Of this, two are at the linear frequencies as given by \eqref{chiI} and \eqref{chiw}
\begin{eqnarray}
\chi^I(x,T) =\sum_{n} \;c_{n{\rm NL}}\; \phi_-(x,s_n) \;e^{Ts_n},
\end{eqnarray}
and
\begin{eqnarray}
\chi^S(x,T)\Big|_{W \text{ poles}} =\sum_{n}c_{W}e^{Ts_n}\phi_{-}(x,s_n).
\end{eqnarray}
There is a third contribution $\chi^{S}\Big|_{S \text{ poles}}$ which gives the contribution from the poles of the Laplace transformed source term.
Now, a source term quadratic in the linear order solutions can be written as
\begin{eqnarray}
    S(x,T) = \sum_{i,j=0, i \leq j}^{n,n} H_{ij}(x)c_{iL}\phi_-(x,s_i)c_{jL}\phi_-(x,s_j)e^{T(s_i+s_j)}.
\end{eqnarray}
Taking the Laplace transform of this gives us
\begin{eqnarray}
    S(x,s) = \sum_{i,j=0, i\leq j}^{n,n}\frac{c_{iL}c_{jL}}{s-s_i-s_j}H_{ij}(x)\phi_-(x,s_i)\phi_-(x,s_j).
\end{eqnarray}
There are simple poles at all $s=(s_i + s_j)$.
Similar to \eqref{chiS}, this gives us
\begin{eqnarray}
 \chi^S(x \rightarrow \infty,T) = \sum_{i,j=0, i\leq j}^{n,n}\frac{c_{iL}c_{jL}e^{(s_i+s_j)T}\phi_+(x,s_i+s_j)}{W(s_i+s_j)}\int_{-\infty}^{\infty}{\rm d}x'\phi_-(x',s_i+s_j)\phi_-(x',s_i)\phi_-(x',s_j)H_{ij}(x'),
 \label{overtone}
\end{eqnarray}
and similarly for the limit $x \rightarrow -\infty$, $\phi_+(x,s_i+s_j)$ goes inside the integral and $\phi_-(x,s_i+s_j)$ comes outside.
We can compute the integral in \ref{overtone} by the method of steepest descent, provided the approximation of stationary phase is valid.

Now, let us consider the general case of two linear $l_1$ modes interacting to give rise to a second order $l_2$ mode. We can define the ratio of the second order amplitude to the product of the two parent first order amplitudes. This ratio helps us quantify the relative amplitude of these QQNMs to the corresponding linear parent modes and is the natural ratio to take since the time dependence in the numerator and denominator gets cancelled out.
Let us define
\begin{eqnarray}
    NL_{ij}(x\rightarrow \infty, s_i+s_j) &=& \left|\frac{1}{s_i+s_j}\frac{c_{iL}c_{jL}e^{(s_i+s_j)T}\phi_+(x \rightarrow \infty,s_i+s_j)}{c_{iL}c_{jL}e^{(s_i+s_j)T}\phi_-(x \rightarrow \infty,s_i)\phi_-(x \rightarrow \infty,s_j)W(s_i+s_j)}\right|\times\nonumber\\
    &&\left|\int_{-\infty}^{\infty}{\rm d}x'\phi_-(x',s_i+s_j)\phi_-(x',s_i)\phi_-(x',s_j)H_{ij}(x')\right|, \nonumber\\
    &=&\left|\frac{1}{s_i+s_j}\sqrt{\frac{s_i s_j}{s_i+s_j}}\frac{1}{2 i \tilde{B}_{ij}}(I_{1ij}+I_{2ij})\right|,
\end{eqnarray}
where $I_{1ij}$ and $I_{2ij}$ are similar to the $l=4$ case. Here, $a_{s}$, $b_{s}$ and $A_{-,s}$ are functions of $s$ as defined in \eqref{phi-qnm},\eqref{phi-}.
\begin{eqnarray}
    I_{1ij} &=& a_{s_i+s_j}A_{-,s_i}A_{-,s_j}H_{ij}(r_{m1})\times\nonumber\\
    &&\left[\frac{2^{\frac{\nu'}{2}\sqrt{\pi}}}{\Gamma(\frac{1-\nu'}{2})}{}_1F_1\left(\frac{-\nu'}{2},\frac{1}{2},\frac{t^2_{l_2}(x_{m1})}{2}\right)-\frac{\sqrt{2\pi}2^{\frac{\nu'}{2}}}{\Gamma(\frac{-\nu'}{2})}t_{l_2}(x_{m1}){}_1F_1\left(\frac{1-\nu'}{2},\frac{3}{2},\frac{t^2_{l_2}(x_{m1})}{2}\right)\right]\times\nonumber\\
    &&\left[\frac{2^{\frac{i}{2}\sqrt{\pi}}}{\Gamma(\frac{1-i}{2})}{}_1F_1\left(\frac{-i}{2},\frac{1}{2},\frac{t^2_{l_1}(x_{m1})}{2}\right)-\frac{\sqrt{2\pi}2^{\frac{i}{2}}}{\Gamma(\frac{-i}{2})}t_{l_1}(x_{m1}){}_1F_1\left(\frac{1-i}{2},\frac{3}{2},\frac{t^2_{l_1}(x_{m1})}{2}\right)\right]\times\nonumber\\
    &&\left[\frac{2^{\frac{j}{2}\sqrt{\pi}}}{\Gamma(\frac{1-j}{2})}{}_1F_1\left(\frac{-j}{2},\frac{1}{2},\frac{t^2_{l_1}(x_{m1})}{2}\right)-\frac{\sqrt{2\pi}2^{\frac{j}{2}}}{\Gamma(\frac{-j}{2})}t_{l_1}(x_{m1}){}_1F_1\left(\frac{1-j}{2},\frac{3}{2},\frac{t^2_{l_1}(x_{m1})}{2}\right)\right]\times\nonumber\\
    &&e^{\frac{-ia' b'(x_{l_1}-x{l_2})^2}{a'+b'}}\sqrt{\frac{\pi}{i(a'+b')}},
\end{eqnarray}
and
\begin{eqnarray}
    I_{2ij} &=& b_{s_i+s_j}A_{-,s_i}A_{-,s_j}H_{ij}(r_{m2})\times\nonumber\\
    &&\left[\frac{2^{\frac{-1-\nu'}{2}\sqrt{\pi}}}{\Gamma(1+\frac{\nu'}{2})}{}_1F_1\left(\frac{1+\nu'}{2},\frac{1}{2},\frac{-t^2_{l_2}(x_{m2})}{2}\right)-\frac{\sqrt{2\pi}2^{\frac{-1-\nu'}{2}}}{\Gamma(\frac{1+\nu'}{2})}it_{l_2}(x_{m2}){}_1F_1\left(1+\frac{\nu'}{2},\frac{3}{2},\frac{-t^2_{l_2}(x_{m1})}{2}\right)\right]\times\nonumber\\
    &&\left[\frac{2^{\frac{i}{2}\sqrt{\pi}}}{\Gamma(\frac{1-i}{2})}{}_1F_1\left(\frac{-i}{2},\frac{1}{2},\frac{t^2_{l_1}(x_{m2})}{2}\right)-\frac{\sqrt{2\pi}2^{\frac{i}{2}}}{\Gamma(\frac{-i}{2})}t_{l_1}(x_{m2}){}_1F_1\left(\frac{1-i}{2},\frac{3}{2},\frac{t^2_{l_1}(x_{m2})}{2}\right)\right]\times\nonumber\\
    &&\left[\frac{2^{\frac{j}{2}\sqrt{\pi}}}{\Gamma(\frac{1-j}{2})}{}_1F_1\left(\frac{-j}{2},\frac{1}{2},\frac{t^2_{l_1}(x_{m2})}{2}\right)-\frac{\sqrt{2\pi}2^{\frac{j}{2}}}{\Gamma(\frac{-j}{2})}t_{l_1}(x_{m2}){}_1F_1\left(\frac{1-j}{2},\frac{3}{2},\frac{t^2_{l_1}(x_{m2})}{2}\right)\right]\times\nonumber\\
    &&e^{\frac{ia' b'(x_{l_1}-x{l_2})^2}{b'-a'}}\sqrt{\frac{\pi}{i(b'-a')}}.
\end{eqnarray}
Here,
\begin{eqnarray}
    \nu' = \nu_{l_2}(s_i+s_j) = i \frac{((s_i+s_j)^2 + V_{l_2}(x_{l_2}))}{\sqrt{-2V_{l_2}''(x_{l_2})}}-\frac{1}{2}, \nonumber\\
     k_{l}=-\frac 12 V_{l}''(x_0),\nonumber\\
    t_{l_2} = (4 k_{l_2})^{\frac{1}{4}}e^{\frac{i\pi}{4}}(x-x_{l_2}), \nonumber\\
    x_{m1} = \frac{a'x_{l_2}+b'x_{l_1}}{a'+b'}, \nonumber\\
    x_{m2} = \frac{b'x_{l_1}-a'x_{l_2}}{b'-a'},\nonumber\\
    a' = \frac{\sqrt{k_{l_2}}}{2},\nonumber\\
    b' = \sqrt{k_{l_1}},\nonumber\\
    x_{l_2} = \text{Maximum point of $l=l_2$ Zerilli potential},\nonumber\\
    x_{l_1} = \text{Maximum point of $l=l_1$ Zerilli potential}.
\end{eqnarray}
Similarly, calculating this ratio at the horizon gives us
\begin{eqnarray}
    NL_{ij}(x\rightarrow -\infty, s_i+s_j) &=& \left|\frac{1}{s_i+s_j}\frac{c_{iL}c_{jL}e^{(s_i+s_j)T}\phi_-(x \rightarrow -\infty,s_i+s_j)}{c_{iL}c_{jL}e^{(s_i+s_j)T}\phi_-(x \rightarrow -\infty,s_i)\phi_-(x \rightarrow -\infty,s_j)W(s_i+s_j)}\right|\times\nonumber\\
    &&\left|\int_{-\infty}^{\infty}{\rm d}x'\phi_+(x',s_i+s_j)\phi_-(x',s_i)\phi_-(x',s_j)H_{ij}(x')\right|,\nonumber\\
    &=& \left|\frac{1}{s_i+s_j}\sqrt{\frac{s_i s_j}{s_i+s_j}}\frac{1}{2 i \tilde{B}_{ij}}\int_{-\infty}^{\infty}{\rm d}x'\phi_+(x',s_i+s_j)\phi_-(x',s_i)\phi_-(x',s_j)H_{ij}(x')\right|.\nonumber \\
\end{eqnarray}
Further following from \eqref{conversion}, to write the nonlinearity in terms of gravitational strain there will be an extra $\frac{4 s_i^2 s_j^2}{(s_i+s_j)^2}$ factor. We now want to take $l=4$, and compute the ratio when the linear parent modes are the $n=1$ and $n=1$ modes and for the case when the parent modes are the $n=0$ and $n=1$ modes respectively.
However, we have checked that the method of steepest descent is not valid for the relevant integrals with the naive choice of contour \ref{contour_of_choice}. We thus perform numerical integration in Mathematica to evaluate the value of the relevant integral between the matching points, taking them to be the turning points at the linear order for $s_0$. This gives us a rough estimate for the nonlinearities (for the gravitational strain) as
\begin{eqnarray}
    NL_{l=4, N-integrate}(2s_1,\infty) = 0.575,
\end{eqnarray}
\begin{eqnarray}
    NL_{l=4, N-integrate}(2s_1,-\infty) = 0.496,
\end{eqnarray}
and
\begin{eqnarray}
    NL_{l=4, N-integrate}(s_0+s_1,\infty) = 0.295,
\end{eqnarray}
\begin{eqnarray}
    NL_{l=4, N-integrate}(s_0+s_1,-\infty) = 0.226.
\end{eqnarray}
The method of steepest descent does turn out to be valid for the case of integral $I_2$ for the $2s_1$ overtone for the contour of our choice, which we can use to give an upper bound on the nonlinearity at infinity with respect to the nonlinearity at horizon. We have,
\begin{eqnarray}
    &&NL(2s_1, \infty) = \left|\frac{1}{2s_1}\sqrt{\frac{s_1}{2}}\frac{1}{W(2s_1)}s_1^2\right|\times \nonumber\\
    &&\left|a_{2s_1}\int_{-\infty}^{\infty}{\rm d}x'D_{\nu_{2s_1}}(t_{l=4}(x'))\phi_-(x',s_1)^2 + b_{2s_1}\int_{-\infty}^{\infty}{\rm d}x'D_{-1-\nu_{2s_1}}(it_{l=4}(x'))\phi_-(x',s_1)^2\right|,\label{2s1 infinity}
\end{eqnarray}
and
\begin{eqnarray}
    &&NL(2s_1, -\infty) = \left|\frac{1}{2s_1}\sqrt{\frac{s_1}{2}}\frac{1}{W(2s_1)}s_1^2\right|\times \left|A_{+,2s_1}\int_{-\infty}^{\infty}{\rm d}x'D_{\nu_{2s_1}}(t_{l=4}(x'))\phi_-(x',s_1)^2 \right|.
\end{eqnarray}
Using the triangle inequality gives us
\begin{eqnarray}
    NL_{l=4}(2s_1, \infty) \leq \left|\frac{a_{2s_1}}{A_{+,2s_1}}\right|NL_{l=4}(2s_1, -\infty) + \left|\frac{1}{2s_1}\sqrt{\frac{s_1}{2}}\frac{1}{W(2s_1)}s_1^2\; b_{2s_1}\int_{-\infty}^{\infty}{\rm d}x'D_{-1-\nu_{2s_1}}(it_{l=4}(x'))\phi_-(x',s_1)^2\right.\nonumber\\
\end{eqnarray}
Evaluating the terms gives us
\begin{eqnarray}
    NL_{l=4}(2s_1, \infty) \leq 0.102 + 0.532 NL_{l=4}(2s_1, -\infty).
\end{eqnarray}

Nonlinearity ratios for overtones in several channels have been obtained using the Leaver algorithm in \cite{Bucciotti_2024}. What is the importance of this calculation? Nonlinearities, such as these ratios, are very sensitive to the particular theory of gravity. By computing these ratios in general relativity, we can compare with observations for direct signatures of corrections to general relativity. We can also potentially compute the changes in these ratios in an effective field theory of gravity that has corrections to general relativity.

\section{Summary and Discussion}\label{sec12}

It has been recognized that when a black hole formed from a merger rings down, one has to go beyond linear perturbation theory and linear QNMs while studying the gravitational wave signal. Quadratic QNMs are excited, and can have sometimes larger amplitudes than linear overtones. These QQNMs are sourced by terms quadratic in the linear perturbation. To know which QQNM to look for in the gravitational wave signal, it is very important to compute the amplitude of the QQNM relative to the square of the amplitude of the linear QNM sourcing it. The ratio of these two quantities at infinity is the nonlinearity ratio. Following \cite{Perrone_2024}, we compute the nonlinearity ratio in a WKB approximation, using matched asymptotic expansions and the method of steepest descent to evaluate a relevant integral.

In \cite{Perrone_2024}, the matching procedure used amounted to considering the QQNM frequency to be one of the linear QNM frequencies, which is generally not the case. Therefore, we do not assume this. We compute the QQNM mode solution sourced by the dominant $(l,m) = (2,2)$ mode, resulting in a $(4,4)$ mode. The nonlinearity ratio we obtain matches very well to that obtained from numerical simulations.

The procedure however, has its limitations, and we explain this for the $(2,0) \times (2,0) \to (2,0)$ channel, which is excited by head-on collisions of non-spinning black holes. Unfortunately, the computation of the nonlinearity ratio relies on the method of steepest descent, which we show is not a good approximation for this channel. However, the simplistic idea that one could numerically evaluate the integral required to compute the nonlinearity ratio also fails here, since considerations of causality imply that the linear QNMs sourcing this QQNM are spatially truncated. We do not yet have a precise region where the spatially truncated QNMs have non-zero support. Therefore, when computing the nonlinearity ratios for a channel in this analytical approximation, we have to first check whether the method of steepest descent can be applied. When it cannot be applied, we can only get a rough estimate of the ratio by numerically integrating between matching points. Other creative methods can be employed, such as in \cite{kehagias2025nonlinearitiesschwarzschildblackhole} for computing the nonlinearity ratio for this channel.

Recently, nonlinear ringdown has also been observed at the horizon \cite{Khera_2023}, and we can compute a similar nonlinearity ratio at the horizon to study this. We have done this for the first time, for the $(2,2) \times (2,2) \to (4,4)$ channel. The study of \cite{Khera_2023} was for head-on collisions, and their modes had $m=0$, therefore direct comparison with their results are not possible; however the ratios are of the same order.

There are several subtleties in the use of second order perturbation theory. The second order perturbation is sourced by terms quadratic in the first order perturbations. However, these and the source depend on the gauge choice at first order. If we use Regge-Wheeler gauge at first order, the source is badly behaved asymptotically as $r \to \infty$. This is because Regge-Wheeler gauge is not asymptotically flat. This can be remedied by a regularization, which amounts to changing to asymptotically flat gauge. However, this regularization is not unique. In the WKB + steepest descent method we use, it is not a priori evident that the nonlinearity ratios will be unchanged for different choices of regularizations that give identical results at infinity. We consider some choices of regularizations which are equivalent as $r \to \infty$, and the nonlinearity ratio stays almost the same, both at the horizon and infinity.

We discuss how to pick points at which matching of the solution across different regions is done. We also have to see if the nonlinearity ratio is sensitive to the points at which we match the solution across various regions, provided the points are in the allowed range. We find that it is not sensitive to this choice.

Finally, we also study QQNMs sourced by overtones. While this may not be observationally as significant as the dominant mode, an analytical approximation scheme such as the one we use, allows us to compute, in principle, the amplitude of QQNMs sourced by different overtones. We can define a ratio of the QQNM amplitude divided by a product of amplitudes of its linear parent modes and compute this. This gives a theoretical idea of which linear modes lead to significant nonlinearities in general relativity. These ratios will change if we have a different theory of gravity, since the structure of the nonlinearities will be different. So by seeing these relative amplitudes in a GW signal, one will be able to infer whether there are corrections to general relativity.

One could study more general channels, for which source terms must be evaluated --- at present, we only have a handful of computations of source terms for various channels even just for the Schwarzschild black hole. We can also compute nonlinearity ratios more precisely, taking into account the renormalization of the linear QNM by various channels.

Another generalization is to study second order perturbations of Kerr black holes, pioneered by Campanelli and Lousto \cite{PhysRevD.59.124022} and studied more recently in \cite{PhysRevD.103.104017}. We aim to compute nonlinearity ratios for various channels in the perturbations of the Kerr black hole. Kerr nonlinearity ratios in the eikonal limit have been computed in \cite{perrone2025nonlinearitieskerrblackhole}, and at the light ring of the Kerr black hole in \cite{JRY2}. Numerical simulations of rotating black hole merger, with a discussion on the QQNMs can be found in \cite{zhu2024nonlineareffectsblackhole}. Kerr QQNMs have also been discussed numerically in \cite{Ma_2024}, \cite{Khera_2025}.

Finally, one will need to develop the analytical approximation to compute the nonlinearity ratios in higher curvature gravity theories (which are typically given by higher than second order equations of motion). These could be completely different from general relativity (see also \cite{Kehagias2024detectdeviationseinsteinsgravity} for an argument against this using causality). Natural generalizations include studying second order perturbation theory in at least a simple higher curvature gravity for a spinless black hole. One could also consider an effective field theory where the higher curvature corrections are small, and do an analysis of QQNMs in perturbation theory in the coupling constants of the higher curvature terms.

We expect analytical approximations to study nonlinear ringdown to be invaluable in computing precision black hole ringdown involving the quadratic quasinormal modes. The one challenge that must be overcome in order to do this within the method discussed QQNM is to find the precise spatial support of the truncated QNMs. Other methods \cite{Zengino_lu_2011}, \cite{Panosso_Macedo_2024} bypassing this problem by hyperboloidal time slicings have been discussed recently in \cite{bourg2025quadraticquasinormalmodesnull}.
\section{Acknowledgments}
JS wishes to acknowledge the support of KVPY and Infosys fellowship during the period 2024-2025.
\section{Appendix}
Here we list all the sources used in the calculation of non-linearity dependence on gauge. The original source is regularized by subtracting from $\chi$
\begin{eqnarray}
     \chi_0 = \frac{1}{126}\sqrt{\frac{70}{\pi}}\frac{(r-2)^2}{r}\psi_{,r}\psi_{,rT}
\end{eqnarray}
We change this term according the gauges used to acquire the new source terms. The source term corresponding to the different gauges are thus as follows:

\subsection{Source term for $\chi_0 = \frac{1}{126}\sqrt{\frac{70}{\pi}}\left[\frac{(r-2)^2}{r}+\frac{16}{r^3}\right]\psi_{,T}\psi_{,TT}$}

\begin{eqnarray}
&\frac{1}{9}\sqrt{\frac{5}{14\pi}}s\bigg[\bigg(-\frac{9 (-2 + r)^2 (-144 - 1341 r - 4182 r^2 - 1296 r^3 + 11630 r^4 +
   31596 r^5 + 41124 r^6 + 30560 r^7 + 11760 r^8 + 2160 r^9)}{(3 + 2 r)^4 (1 + 3 r)^2(-2 + r) r^7}+\nonumber\\
&\frac{3 r^4 (-96 - 555 r - 800 r^2 - 816 r^3 - 1389 r^4 - 1470 r^5 +
   476 r^6 + 276 r^7) s^2}{(-3 + 7 r + 6 r^2)^2(-2 + r) r^7}+\frac{r^8 (4 + 7 r) s^4}{(-2 + r) r^7}\bigg)\psi^2-\nonumber\\
&\frac{2 (-2 + r)\bigg(\frac{3 (-144 - 1119 r - 3540 r^2 - 9512 r^3 - 8230 r^4 - 2759 r^5 +
   2154 r^6 + 4116 r^7 + 144 r^8)}{(3 + 2 r)^3}+2 (-2 + r) r^4 s^2\bigg)\psi \psi_r}{r^6 (1 + 3 r)^2} -\nonumber\\
&\frac{(-2 + r)\bigg(-\frac{3 (-48 - 273 r - 514 r^2 - 384 r^3 + 142 r^4 - 370 r^5 + 8 r^6 +
   228 r^7)}{(3 + 2 r)^2 (1 + 3 r)^2}+r^4 (4 + 7 r) s^2\bigg)\psi_r^2}{r^5}-\nonumber\\
&\frac{2 (-2 + r)^2}{r^5 (1 + 3 r)^2}\bigg((-8 + 3 r (-18 + r (-35 + 6 r (-9 + 5 r))) + 2 r^4 (s + 3 r s)^2)\psi_r^2-r^2 (2 + (5 - 3 r) r)^2\psi_{rr}^2-\nonumber\\
&(-2 + r) r (1 + 3 r)^2\psi_r(2 (3 + r) \psi_{rr} + (-2 + r) r
\psi_{rrr}\bigg)\nonumber\\
&+\bigg(4s^2+2\frac{(r - 2)(90 r^3 + 27 r^2 + 9 r + 1)}{r^4 (3 r + 1)^2}\bigg)\bigg(\left[\frac{(r-2)^2}{r}+\frac{16}{r^3}\right]s^3\psi\psi\bigg)-\frac{(r-2)^2}{r^2}\partial_{rr}\bigg(\left[\frac{(r-2)^2}{r}+\frac{16}{r^3}\right]s^3\psi\psi\bigg)\nonumber\\
&-\frac{2(r-2)}{r^3}\partial_r\bigg(\left[\frac{(r-2)^2}{r}+\frac{16}{r^3}\right]s^3\psi\psi\bigg)
\bigg]
\end{eqnarray}

\subsection{Source term for $\chi_0 = \frac{1}{126}\sqrt{\frac{70}{\pi}}\left[\frac{(r-2)^2}{r}+\frac{16}{r^3}\right]\psi_{,x}\psi_{,xT}$}

\begin{eqnarray}
&\frac{1}{9}\sqrt{\frac{5}{14\pi}}s\bigg[\bigg(-\frac{9 (-2 + r)^2 (-144 - 1341 r - 4182 r^2 - 1296 r^3 + 11630 r^4 +
   31596 r^5 + 41124 r^6 + 30560 r^7 + 11760 r^8 + 2160 r^9)}{(3 + 2 r)^4 (1 + 3 r)^2(-2 + r) r^7}+\nonumber\\
&\frac{3 r^4 (-96 - 555 r - 800 r^2 - 816 r^3 - 1389 r^4 - 1470 r^5 +
   476 r^6 + 276 r^7) s^2}{(-3 + 7 r + 6 r^2)^2(-2 + r) r^7}+\frac{r^8 (4 + 7 r) s^4}{(-2 + r) r^7}\bigg)\psi^2-\nonumber\\
&\frac{2 (-2 + r)\bigg(\frac{3 (-144 - 1119 r - 3540 r^2 - 9512 r^3 - 8230 r^4 - 2759 r^5 +
   2154 r^6 + 4116 r^7 + 144 r^8)}{(3 + 2 r)^3}+2 (-2 + r) r^4 s^2\bigg)\psi \psi_r}{r^6 (1 + 3 r)^2} -\nonumber\\
&\frac{(-2 + r)\bigg(-\frac{3 (-48 - 273 r - 514 r^2 - 384 r^3 + 142 r^4 - 370 r^5 + 8 r^6 +
   228 r^7)}{(3 + 2 r)^2 (1 + 3 r)^2}+r^4 (4 + 7 r) s^2\bigg)\psi_r^2}{r^5}-\nonumber\\
&\frac{2 (-2 + r)^2}{r^5 (1 + 3 r)^2}\bigg((-8 + 3 r (-18 + r (-35 + 6 r (-9 + 5 r))) + 2 r^4 (s + 3 r s)^2)\psi_r^2-r^2 (2 + (5 - 3 r) r)^2\psi_{rr}^2-\nonumber\\
&(-2 + r) r (1 + 3 r)^2\psi_r(2 (3 + r) \psi_{rr} + (-2 + r) r
\psi_{rrr}\bigg)\nonumber\\
&+\bigg(4s^2+2\frac{(r - 2)(90 r^3 + 27 r^2 + 9 r + 1)}{r^4 (3 r + 1)^2}\bigg)\bigg(\left[\frac{(r-2)^2}{r}+\frac{16}{r^3}\right]\psi_{,x}\psi_{,xT}\bigg)-\frac{(r-2)^2}{r^2}\partial_{rr}\bigg(\left[\frac{(r-2)^2}{r}+\frac{16}{r^3}\right]\psi_{,x}\psi_{,xT}\bigg)\nonumber\\
&-\frac{2(r-2)}{r^3}\partial_r\bigg(\left[\frac{(r-2)^2}{r}+\frac{16}{r^3}\right]\psi_{,x}\psi_{,xT}\bigg)
\bigg]
\end{eqnarray}

\subsection{Source term for $\chi_0 = \frac{1}{126}\sqrt{\frac{70}{\pi}}\left[\frac{(r-2)^2}{r}+\frac{16}{r^3}\right]\psi_{,x}\psi_{,xT}$}

\begin{eqnarray}
&\frac{1}{9}\sqrt{\frac{5}{14\pi}}s\bigg[\bigg(-\frac{9 (-2 + r)^2 (-144 - 1341 r - 4182 r^2 - 1296 r^3 + 11630 r^4 +
   31596 r^5 + 41124 r^6 + 30560 r^7 + 11760 r^8 + 2160 r^9)}{(3 + 2 r)^4 (1 + 3 r)^2(-2 + r) r^7}+\nonumber\\
&\frac{3 r^4 (-96 - 555 r - 800 r^2 - 816 r^3 - 1389 r^4 - 1470 r^5 +
   476 r^6 + 276 r^7) s^2}{(-3 + 7 r + 6 r^2)^2(-2 + r) r^7}+\frac{r^8 (4 + 7 r) s^4}{(-2 + r) r^7}\bigg)\psi^2-\nonumber\\
&\frac{2 (-2 + r)\bigg(\frac{3 (-144 - 1119 r - 3540 r^2 - 9512 r^3 - 8230 r^4 - 2759 r^5 +
   2154 r^6 + 4116 r^7 + 144 r^8)}{(3 + 2 r)^3}+2 (-2 + r) r^4 s^2\bigg)\psi \psi_r}{r^6 (1 + 3 r)^2} -\nonumber\\
&\frac{(-2 + r)\bigg(-\frac{3 (-48 - 273 r - 514 r^2 - 384 r^3 + 142 r^4 - 370 r^5 + 8 r^6 +
   228 r^7)}{(3 + 2 r)^2 (1 + 3 r)^2}+r^4 (4 + 7 r) s^2\bigg)\psi_r^2}{r^5}-\nonumber\\
&\frac{2 (-2 + r)^2}{r^5 (1 + 3 r)^2}\bigg((-8 + 3 r (-18 + r (-35 + 6 r (-9 + 5 r))) + 2 r^4 (s + 3 r s)^2)\psi_r^2-r^2 (2 + (5 - 3 r) r)^2\psi_{rr}^2-\nonumber\\
&(-2 + r) r (1 + 3 r)^2\psi_r(2 (3 + r) \psi_{rr} + (-2 + r) r
\psi_{rrr}\bigg)\nonumber\\
&+\bigg(4s^2+2\frac{(r - 2)(90 r^3 + 27 r^2 + 9 r + 1)}{r^4 (3 r + 1)^2}\bigg)\bigg(\left[\frac{(r-2)^2}{r}+\frac{16}{r^3}\right]s \psi_{,x}^2\bigg)-\frac{(r-2)^2}{r^2}\partial_{rr}\bigg(\left[\frac{(r-2)^2}{r}+\frac{16}{r^3}\right]s \psi_{,x}^2\bigg)\nonumber\\
&-\frac{2(r-2)}{r^3}\partial_r\bigg(\left[\frac{(r-2)^2}{r}+\frac{16}{r^3}\right]s \psi_{,x}^2\bigg)
\bigg]
\end{eqnarray}

\end{document}